\documentclass[%
 reprint,
superscriptaddress,
 amsmath,amssymb,
 aps,
pra,
]{revtex4-2}

\usepackage{graphicx}
\usepackage{dcolumn}
\usepackage{bm}

\begin{document}


\title{Amplification of genuine tripartite nonlocality and entanglement in the Schwarzschild spacetime under decoherence}

\author{Chunyao Liu}
\affiliation{%
 College of Physics, Guizhou University, Guiyang 550025, China
}%
\affiliation{%
School of Physics and Electronics, Guizhou Normal University, Guiyang 550001, China
}%
 
\author{Zhengwen Long}%
 \email{zwlong@gzu.edu.cn (corresponding author)}
\affiliation{%
College of Physics, Guizhou University, Guiyang 550025, China
}%


\author{Qiliang He}
 \email{heliang005@163.com}
\affiliation{%
School of Physics and Electronics, Guizhou Normal University, Guiyang 550001, China
}%


\date{\today}

\begin{abstract}
We investigate the amplification of the genuine tripartite nonlocality(GTN) and the genuine tripartite entanglement(GTE) of Dirac particles in the background of a Schwarzschild black hole by a local filtering operation under decoherence. It is shown that the physically accessible GTN will be completely destroyed by decoherence, which means that the physically accessible GTN will not exist in the system. Particularly, the local filtering operation can make the physically accessible GTN appear within a certain range of Hawking temperature, namely, the local filtering operation can cause the physically accessible GTN to be generated in the system coupled with the environment, which is not discovered before and is benefit for the quantum information processing. Furthermore, we also find that the physically accessible GTE approaches a stable value in the limit of infinite Hawking temperature for most cases, but if the decoherence parameter $p$ is less than 1, the ``sudden death'' of GTE will take place when the decoherence strength is large enough. It is worth noting that the nonzero stable value of GTE can be increased by performing the local filtering operation, even in the presence of decoherence. Finally, we explore the generation of physically inaccessible GTN and GTE of other tripartite subsystems under decoherence, it is shown that the physically inaccessible GTN cannot be produced, but the physically inaccessible GTE can be produced. In addition, we can see that the generated physically inaccessible GTE can be increased by applying the local filtering operation.
\end{abstract}

\maketitle


\section{\label{sec:level1}Introduction}

Quantum nonlocality, which originated in Einstein's discussion in 1935 \cite{a1a}, has attracted widespread attention in quantum mechanics and quantum information science research \cite{a2a,a3a,a4a,a5a,a6a}. Many studies have shown that quantum nonlocality, as a quantum resource, plays a very important role in various quantum information tasks \cite{a7a,a8a,a9a,a10a,a11a,a12a}, and has become a hot topic of intense research in recent years. In 1987, Svetlichny introduced the so-called Svetlichny inequality \cite{a13a}, which can be used as a measure for quantifying the GTN of three-body systems. Some studies have proven that being GTN is a sufficient, but not always necessary, condition for being GTE. Therefore, the study of many-body non-locality not only helps us to further understand the quantum world but also opens up new applications in quantum communication and quantum computing.

On the other hand, relativistic quantum information, a combination of relativity theory and quantum information science, has attracted a lot of attention \cite{a14a,a15a,a16a,a18a,a19a} in recent years. The study of the influence of relativistic effects on quantum correlation dynamics and quantum information processing is not only helpful in understanding some key questions in quantum information theory \cite{a20a,a22a,a21a} but also gives a new method of understanding the information paradox existing in black holes \cite{a23a,a24a,a26a,a27a,a25a}. Following the work of pioneers, many efforts have been devoted to investigating quantum information theory in the relativistic framework\cite{a29a,a30a,a32a,a34a,a33a,aa33aa,aa34aa}. For instance, Wang et al. \cite{a29a} derived the quantum discord in noninertial frames. Yao et al. \cite{a30a} investigated the quantum Fisher information under the Unruh effect. Wang et al. \cite{a32a} discuss the degree of multipartite entanglement will be degraded due to the Unruh effect in noninertial frames. Wu et al. \cite{a34a} and Zhang et al. \cite{a33a} expanded the investigation of the effect of Hawking radiation on tripartite nonlocality and entanglement in the Schwarzschild black hole, respectively. Li et al. \cite{aa33aa}
expanded the investigation in the background of the Garfinkle-Horowitz-Strominger (GHS) dilation spacetime.

However, the above investigations, such as Refs. \cite{a34a,a33a,aa33aa}, are confined to isolated systems used for quantum information research and do not discuss the impact of the environment. In reality, a real quantum system will unavoidably be influenced by surrounding environments, that is, the real quantum system inevitably suffers from quantum decoherence. Decoherence can be considered as the process of information transferred from the system to the environment\cite{a35a,a36a}, and
is the main challenge for the realization of quantum information processing. Some researchers have devoted themselves to studying the influence of decoherence environments on the quantum entanglement\cite{a37a,a38a,a39a,a40a} in
noninertial frames over recent years. In particular, Haddadi et al.\cite{aa40aa} studied the impacts of Hawking radiation and decoherence on quantum systems near the event horizon of Schwarzschild black holes and found some impressive results, which not only enrich our understanding of quantum correlation degrades in the vicinity of black holes but also provide assistance for the efficient implementation of relativistic quantum information tasks under decoherence. In addition, some authors have searched the ways to overcome decoherence in non-relativistic systems\cite{a42a,a43a,a44a,a46a,a47a} or noninertial frames\cite{a48a,aa48aa}. However, litter studies focus on enhancing or recovering the damaged GTN and GTE for open Dirac system in Schwarzschild spacetime. Therefore, it is necessary to understand the influence of decoherence environments on the GTN and GTE in Schwarzschild spacetime and find efficient ways to prevent or minimize the influence of decoherence.

\begin{figure}
\begin{center}
{\includegraphics[width=7cm]{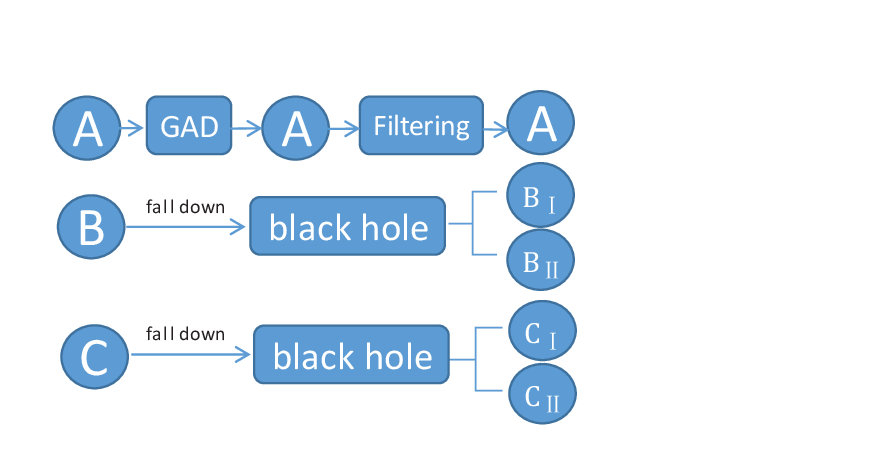}}
\caption{\label{fig:fig1}  The physical scheme diagram which is investigated in the present paper: Alice, Bob, and Charlie, initially share a generically tripartite entangled state at the same initial point in flat Minkowski space-time. Then, Bob and Charlie freely fall into the Schwarzschild black hole. In addition, Alice suffers from GAD noise and then makes a local filtering operation. }
\end{center}
\end{figure}

In this paper, we extend the previous research \cite{a34a} from an isolated system to an open system, namely, we study the effect of environmental decoherence on the GTN and GTE of the system in the background of a Schwarzschild black hole and focus on how to enhance the GTN and GTE of the system, the detailed physical schematic diagram as shown in Fig.1. We assume there are three observers, Alice, Bob, and Charlie, initially sharing a generically tripartite entangled state at the same initial point in flat Minkowski space-time. Then, Bob and Charlie freely fall into the Schwarzschild black hole and are finally located near the event horizon. In addition, Alice suffers from decoherence noise and makes a filtering operation. The main end of this paper is to study the influence of decoherence on the physically accessible GTN and GTE, the production of the physically inaccessible GTN and GTE of Dirac particles in the background of a Schwarzschild black hole, and explore the amplification of the GTN and GTE of the open Dirac system by the local filtering operation.

The paper is organized as follows. In Sec. 2, we briefly recall the measures of GTN and GTE for tripartite quantum systems. In section 3, the quantization of Dirac fields in a Schwarzschild black hole is reviewed. In section 4, we discuss the influence of decoherence on the dynamics of the physically accessible GTN and GTE, as well as the physically inaccessible GTN and GTE. In addition, we present a physical scheme for enhancing the GTN and GTE of the system. Finally, Section 5 is the conclusions and discussions.

\section{A BRIEF REVIEW OF GTN AND GTE}
In this section, let us briefly review the concepts of GTN and GTE. In tripartite systems, nonlocality can be used to demonstrate genuine tripartite correlations. In general, the local tripartite correlations shared by Alice, Bob, and Charlie can be described as
\begin{equation}
   P(a,b,c\mid x,y,z) =\sum_{\lambda} q_{\lambda }P_{\lambda }(a\mid x) P_{\lambda }(b\mid y) P_{\lambda }(c\mid z),
\end{equation}
where $0\le q_{\lambda }\le1$, $\sum_{\lambda}q_{\lambda }=1$. $x,y,z\in\left \{0,1\right \}$ and $a,b,c\in\left \{0,1\right \}$ are the inputs and outputs of the parties Alice, Bob, and Charlie, respectively. $P_{\lambda }(a\mid x) $ is the
conditional probability for obtaining the output $a$ when the measurement setting is $x$ and $\lambda $ is the hidden state. When the tripartite correlations cannot be written as Eq.(1), it can be said that the tripartite system has nonlocality.

In 1987, Svetlichny introduced a hybrid local-nonlocal form of correlation to detect GTN\cite{a13a}. Under the definition of Svetlichny, a tripartite correlation is called genuine nonlocal if it admits the following local LHV model
\begin{eqnarray}
   P(a,b,c\mid x,y,z) &=& \sum_{\lambda } q_{\lambda }P_{\lambda }(a\mid x) P_{\lambda }(b,c\mid y,z)\nonumber\\
   & +&\sum_{\mu } q_{\mu }P_{\mu } (b\mid y) P_{\mu }(a,c\mid x,z)\nonumber\\
   &+&\sum_{\vartheta } q_{\vartheta }P_{\vartheta }(c\mid z) P_{\vartheta }(a,b\mid x,y),
\end{eqnarray}
where $\sum_{\lambda }q_{\lambda }+      \sum_{\mu } q_{\mu }+\sum_{\vartheta } q_{\vartheta }=1$ and $0\le q_{\lambda }$, $q_{\mu }$, $q_{\vartheta }\le 1$.This kind of correlation is regarded as Svetlichny local, otherwise, they are Svetlichny nonlocal\cite{a49a,a50a,a51a}.
Consider a three-qubit quantum system shared by Alice, Bob, and Charlie. For Alice’s system, the form of measurement observables are $A=a\cdot  \sigma$ and ${A}'={a}'\cdot  \sigma$, where $a=(a_{1}, a_{2}, a_{3})$, ${a}'=({a}'_{1},{a}'_{2},{a}'_{3})\in R^{3}$ are any three-dimensional unit vectors, and $\sigma=(\sigma_{1},\sigma_{2},\sigma_{3})$ is the vector of Pauli matrices. Similarly, we have measurement observables $B$, ${B}'$ and $C$, ${C}'$ for Bob’s system and Clarlie’s system. The Svetlichny operators can be defined as
\begin{eqnarray}
   S &=&(A+{A}') \otimes(B\otimes{C}'+{B}'\otimes C)\nonumber\\&+& (A-{A}') \otimes (B\otimes C-{B}'\otimes{C}').  
\end{eqnarray}

For any three-qubit state $\rho$ which admits the LHV model, the Svetlichny inequality is
\begin{equation}
   tr(S\rho)\le 4
\end{equation}
with $S$ is taken over all possible Svetlichny operators. Equivalently, if a tripartite state violates the Svetlichny inequality for some Svetlichny operator, then it is genuinely nonlocal. For convenience, according to this sufficient condition of genuinely nonlocal, the maximal violation of Svetlichny inequality $S(\rho )$ is  \cite{a52a,a53a}.
\begin{equation}
S(\rho )\equiv \max_{S} tr(S\rho )
\end{equation}
can be used to quantify GTN.

If the density operator of a three-qubit system is a three-qubit $X$ state
\begin{eqnarray}
\rho _{X}= \left( \begin{array}{cccccccc}
 \mu_{1} & 0 & 0 &0& 0 & 0& 0 & w_{1} \\
  0& \mu_{2} & 0 & 0& 0 & 0& w_{2} & 0\\
  0& 0 & \mu_{3} & 0& 0 & w_{3}& 0 & 0\\
 0 & 0 &0  & \mu_{4}& w_{4} & 0& 0 & 0\\
 0 & 0 &0  &w^{*}_{4}& \nu_{4} & 0& 0 & 0\\
 0 & 0 &w^{*}_{3}  &0& 0 & \nu_{3}& 0 & 0\\
 0 & w^{*}_{2} &0  &0& 0 & 0& \nu_{2} & 0\\
 w^{*}_{1} & 0 &0  &0& 0 & 0& 0 & \nu_{1}
\end{array} \right),
\end{eqnarray}
the Svetlichny value can be simply given by \cite{a52a}
\begin{equation}
   S(\rho_{X})=\max \left \{ 8 \sqrt{2} \left | w_{i} \right |,4 \left |  N\right |  \right \} ,
\end{equation}
where $N=\mu_{1}-\mu_{2}-\mu_{3}+\mu_{4}-\nu_{4}+\nu_{3}+\nu_{2}-\nu_{1}$.

Next, let us briefly review some notions of GTE. For a three-partite pure state $\left | \psi  \right \rangle$, if it can be expressed as the form $\left | \psi  \right \rangle =\left | \psi _{A} \right \rangle \otimes \left | \psi _{B} \right \rangle$, where $\left | \psi _{A} \right \rangle$ and $\left | \psi _{B} \right \rangle$ are the monomeric or bipartite pure states, we can say the state $\left | \psi  \right \rangle$ is separable. In addition, if a three-partite state is not separable, then it is called genuinely three-partite entangled. To describe the degrees of the
GTE, the genuine tripartite concurrence of a three-partite pure state $\left | \psi  \right \rangle $ is proposed and defined as
\begin{equation}
 C(\left | \psi  \right \rangle )=\min_{\chi_{i} \in \chi} \sqrt{2 [1-Tr(\rho^{2}_{A_{\chi_{i}}})]}, 
\end{equation}
with $\chi =\left \{ A_{i}\mid B_{i}\right \} $ is the set of all possible bipartitions of the tripartite system, $\rho_{A_{\chi_{i}}}$ denotes the reduced density operator of system $A$ corresponding to the bipartition ${\chi_{i}}$ \cite{a54a,a55a}. Then, the GTE of a mixed state $\rho$ can be described by a convex roof construction
\begin{eqnarray}
C(\rho)=\mathop{\verb|inf|} \limits_{\left \{ p_{i},\left | \psi_{i}  \right \rangle \right \} } \sum_{i} p_{i} C (\left | \psi _{i} \right \rangle ),
\end{eqnarray}
where the infimum is taken over all possible decompositions
$\rho= {\textstyle \sum_{i}} p_{i} |\psi_{i}   \rangle \langle \psi_{i}  | $. In particular, if the density matrix of a three-qubit system has an $X$ form which is given by Eq.(6), the GTE can be written as  
\begin{equation}
C(\rho_{X})=2 \max \left \{0,\left | c_{i} \right | -\vartheta _{i}  \right \} ,i=1,...,4,
\end{equation}
with $\vartheta _{i}= {\textstyle \sum_{j\ne i}^{4}} \sqrt{n_{j} m_{j}} $ \cite{a56a}.

\section{QUANTIZATION OF DIRAC FIELDS IN A SCHWARZSCHILD BLACK HOLE}

The spherically symmetric line element of the Schwarzschild spacetime can be given by 
\begin{equation}
   {\mathrm{d}}s^2=-(1-\frac{2 M}{r}) {\mathrm{d}}t^2+(1-\frac{2 M}{r})^{-1} {\mathrm{d}}r^2+r^2 ({\mathrm{d}}\theta ^2+\sin^{2} \theta \mathrm{d} \psi ^{2} ),
\end{equation}
where $M$ denotes the mass of the black hole.

Solving the Dirac equation $[\gamma ^{a} e^{\mu }_{a } (\partial \mu +\Gamma _{\mu} )] \Psi =0$ \cite{a57a} near the event horizon, the positive (fermions) frequency outgoing solutions outside and inside regions of the event horizon $r_{+}$ can be written as 
\begin{eqnarray}
   \psi _{k}^{\uppercase\expandafter{\romannumeral1} +} =\xi e^{- i \omega u }(r>r_{+}),\nonumber\\
\psi _{k}^{\uppercase\expandafter{\romannumeral2} +} =\xi e^{ i \omega u }(r>r_{+}),     
\end{eqnarray}
where $k$ is the wave vector which labels the modes, $\xi$ represents four-component Dirac spinor, $\omega$ denotes the monochromatic frequency of the Dirac field, $u=t-r_{*}$ and $r_{*}=r+2 M \ln{\frac{r-2M}{2M}}$ is the tortoise coordinate. Particles and antiparticles will be classified by the future-directed time-like Killing vector in each region. 

According to Damour–Ruffini’s suggestion \cite{a58a}, a complete basis of positive energy modes can be obtained by making an analytic continuation for Eq.(12). Afterwards, we can get the Bogoliubov transformations \cite{a59a} between the creation and annihilation operators in the Schwarzschild and Kruskal coordinates through quantizing the Dirac fields in the Schwarzschild and Kruskal modes, respectively. After properly normalizing the state vector, the vacuum state and excited state of the Kruskal particle for mode $k$ can be written as 
\begin{eqnarray}
&&\left | 0 \right \rangle _{k}
=(e^{-\frac{\omega _{k}}{T}}+1)^{-\frac{1}{2}}
\left | 0_{k}  \right \rangle ^{+}_{\uppercase\expandafter{\romannumeral1}}
\left | 0_{-k}  \right \rangle ^{-}_{\uppercase\expandafter{\romannumeral2}}\nonumber\\
&&+
(e^{\frac{\omega _{k}}{T}}+1)^{-\frac{1}{2}}
\left | 1_{k}  \right \rangle ^{+}_{\uppercase\expandafter{\romannumeral1}}
\left | 1_{-k}  \right \rangle ^{-}_{\uppercase\expandafter{\romannumeral2}},\nonumber\\
&&\left | 1 \right \rangle _{k}
=\left | 1_{k}  \right \rangle ^{+}_{\uppercase\expandafter{\romannumeral1}}
\left | 0_{-k}  \right \rangle ^{-}_{\uppercase\expandafter{\romannumeral2}},
\end{eqnarray}
where $T=\frac{1}{8 \pi M}$ is the Hawking temperature \cite{a60a}, $\omega _{k}$ is the frequency of the particles for the mode $k$, $ \left | n_{k} \right \rangle ^{+}_{\uppercase\expandafter{\romannumeral1}} $ and $ \left | n_{-k} \right \rangle ^{-}_{\uppercase\expandafter{\romannumeral2}} $ correspond to the orthonormal bases for the outside and inside regions of the event horizon, respectively. For simplicity, we  consider ${\left | n \right \rangle _{\uppercase\expandafter{\romannumeral1}}}$ and ${\left | n \right \rangle_{\uppercase\expandafter{\romannumeral2}}}$ instead of $ \left | n_{k} \right \rangle ^{+}_{\uppercase\expandafter{\romannumeral1}} $ and $ \left | n_{-k} \right \rangle ^{-}_{\uppercase\expandafter{\romannumeral2}} $ .

\section{THE DYNAMICS OF GTN AND GTE IN SCHWARZSCHILD BLACK HOLE UNDER DECOHERENCE WITH LOCAL FILTERING OPERATION}

Initially, we consider that Alice, Bob, and Charlie share a Greenberger–Horne–Zeilinger-like (GHZ-like) state of the Dirac fields at the same initial point in flat Minkowski spacetime which can be written as 
\begin{equation}
   \left | \Psi   \right \rangle _{A B C}=\alpha \left | 0\right \rangle_{A} \left | 0\right \rangle_{B} \left | 0\right \rangle_{C}+\sqrt{1-\alpha^{2}} \left | 1 \right \rangle_{A} \left | 1 \right \rangle_{B} \left | 1 \right \rangle_{C},  
\end{equation}
where $\alpha$ is the state parameter and $\alpha \in [0,1]$. Next, we assume that Alice stays stationary at the asymptotically flat region, while Bob and Charlie freely fall in toward a Schwarzschild black hole and hover outside the event horizon. We choose the Minkowski mode for Alice and the black hole mode for Bob and Charlie. Using Eq.~(13), we can rewrite Eq. (14) as 
\begin{eqnarray}
& &\left |\Psi \right \rangle_{A B_{\uppercase\expandafter{\romannumeral1}} B_{\uppercase\expandafter{\romannumeral2}} C_{\uppercase\expandafter{\romannumeral1}} C_{\uppercase\expandafter{\romannumeral2}}}
\nonumber\\&&=\alpha (e^{\frac{\omega_{k} }{T} }+1)^{-1} \left | 0  \right \rangle _{A} \left | 1  \right \rangle _{B_{\uppercase\expandafter{\romannumeral1}}} \left | 1  \right \rangle _{B_{\uppercase\expandafter{\romannumeral2}}}
 \left | 1  \right \rangle _{C_{\uppercase\expandafter{\romannumeral1}}} \left | 1  \right \rangle _{C_{\uppercase\expandafter{\romannumeral2}}}\nonumber\\&&+\sqrt{1-\alpha ^{2}} \left | 1  \right \rangle _{A} \left | 1  \right \rangle _{B_{\uppercase\expandafter{\romannumeral1}}} \left | 0  \right \rangle _{B_{\uppercase\expandafter{\romannumeral2}}} \left | 1  \right \rangle _{C_{\uppercase\expandafter{\romannumeral1}}} \left | 0  \right \rangle _{C_{\uppercase\expandafter{\romannumeral2}}}\nonumber\\
 &&+\alpha (e^{\frac{\omega_{k} }{T}} + e^{-\frac{\omega_{k} }{T}}+2)^{-\frac{1}{2}} (
 \left | 0  \right \rangle _{A} \left | 0  \right \rangle _{B_{\uppercase\expandafter{\romannumeral1}}} \left | 0  \right \rangle _{B_{\uppercase\expandafter{\romannumeral2}}} \left | 1  \right \rangle _{C_{\uppercase\expandafter{\romannumeral1}}} \left | 1  \right \rangle _{C_{\uppercase\expandafter{\romannumeral2}}}\nonumber\\&&+ \left | 0  \right \rangle _{A} \left | 1  \right \rangle _{B_{\uppercase\expandafter{\romannumeral1}}} \left | 1  \right \rangle _{B_{\uppercase\expandafter{\romannumeral2}}} \left | 0  \right \rangle _{C_{\uppercase\expandafter{\romannumeral1}}} \left | 0  \right \rangle _{C_{\uppercase\expandafter{\romannumeral2}}})\nonumber\\
&&+\alpha (e^{-\frac{\omega_{k} }{T} }+1)^{-1} \left | 0  \right \rangle _{A} \left | 0  \right \rangle _{B_{\uppercase\expandafter{\romannumeral1}} }\left | 0  \right \rangle _{B_{\uppercase\expandafter{\romannumeral2}}} \left | 0  \right \rangle _{C_{\uppercase\expandafter{\romannumeral1}} }\left | 0  \right \rangle _{C_{\uppercase\expandafter{\romannumeral2}} } .
\end{eqnarray}

Here, we suppose that Alice’s detector sensitive only to mode $\left | n  \right \rangle_{A}$, Bob's and Charlie's detectors outside the event horizon of black hole observe the modes  $\left | n  \right \rangle _{B_{\uppercase\expandafter{\romannumeral1}} }$ and $\left | n  \right \rangle _{C_{\uppercase\expandafter{\romannumeral1}} }$, the modes  $\left | n  \right \rangle _{B_{\uppercase\expandafter{\romannumeral2}} }$  and $\left | n  \right \rangle _{C_{\uppercase\expandafter{\romannumeral2}} }$  are observed by anti-Bob and anti-Charlie inside the event horizon. 

In the following, we consider Alice first undergoes a generalized amplitude damping (GAD) channel, which can be described as\cite{a61a}
\begin{equation}
 \varepsilon _{GAD} (\rho )=\sum_{i=0}^{3} E_{i} \rho E_{i}^{\dagger },
\end{equation}
with 
\begin{eqnarray}
&&E_{0}=\sqrt{p} 
  \left( \begin{array}{cc}
  1 & 0 \\
 0 &\sqrt{1-r} 
\end{array} \right), E_{1}=\sqrt{p}  \left( \begin{array}{cc}
  0&\sqrt{r} \\
 0 &0 
\end{array} \right), \nonumber\\ 
&&E_{2}=\sqrt{1-p}  
\left( \begin{array}{cc}
  \sqrt{1-r}&0 \\
 0 &1 
\end{array} \right), \nonumber\\ &&E_{3}=\sqrt{1-p}  \left( \begin{array}{cc}
  0&0 \\
 \sqrt{r} &0 
\end{array} \right),
\end{eqnarray}
where $r$ is related to the energy relaxation rate and $\left \{ p,r \right \} $ is usually a function of environment temperature ${T}'$, can be read as
\begin{eqnarray}
&r&=1-e^{-\gamma t}(\gamma =[\frac{2}{\exp (-\frac{h \omega }{k_B {T}' } )-1}+1 ] \gamma _{0}), \nonumber\\ 
&p&=\frac{1}{1+{\exp (-\frac{h \omega }{k_B {T}' }) }} .
\end{eqnarray}
with $\gamma_{0}$ denoted the energy relaxation rate, $t$ is the storage period, $h \omega$ and $k_{B}$ are the transition energy of the quantum system and the Boltzmann constant, respectively. Note that setting $p=1$ would reduce the GAD channel to the well-informed amplitude damping (AD) channel. 

Subsequently, Alice makes a filtering operation which can be described as \cite{a62a} 
\begin{eqnarray}
  M_{f t}=
  \left( \begin{array}{cc}
  \sqrt{1-f} & 0 \\
 0 &\sqrt{f} 
\end{array} \right), 0<f<1,
\end{eqnarray}
where $f$ is the strength of the filtering operation. Filtering is a non-trace-preserving map, which is known to be capable of increasing entanglement with some probability \cite{a62a}. With the help of Eqs.~(15), (16), and (19), the total evolution of the system can be written as
\begin{eqnarray}   
{\rho}''_{A B_{\uppercase\expandafter{\romannumeral1}} B_{\uppercase\expandafter{\romannumeral2}} C_{\uppercase\expandafter{\romannumeral1}} C_{\uppercase\expandafter{\romannumeral2}}}=&&\frac{1}{Z} ((M_{ft} \otimes {\uppercase\expandafter{\romannumeral1}_{2}}\otimes {\uppercase\expandafter{\romannumeral1}_{2}}\otimes {\uppercase\expandafter{\romannumeral1}_{2}}\otimes {\uppercase\expandafter{\romannumeral1}_{2}})\nonumber\\ &&\cdot {\rho}'_{A B_{\uppercase\expandafter{\romannumeral1}} B_{\uppercase\expandafter{\romannumeral2}} C_{\uppercase\expandafter{\romannumeral1}} C_{\uppercase\expandafter{\romannumeral2}}}\nonumber\\ &&\cdot(M_{ft} \otimes {\uppercase\expandafter{\romannumeral1}_{2}}\otimes {\uppercase\expandafter{\romannumeral1}_{2}}\otimes {\uppercase\expandafter{\romannumeral1}_{2}}\otimes {\uppercase\expandafter{\romannumeral1}_{2}})^{\dagger }),
\end{eqnarray}
with
\begin{eqnarray}   
{\rho}'_{A B_{\uppercase\expandafter{\romannumeral1}} B_{\uppercase\expandafter{\romannumeral2}} C_{\uppercase\expandafter{\romannumeral1}} C_{\uppercase\expandafter{\romannumeral2}}}=&& {\textstyle \sum_{i=0}^{3}}  (E_{i} \otimes {\uppercase\expandafter{\romannumeral1}_{2}}\otimes {\uppercase\expandafter{\romannumeral1}_{2}}\otimes {\uppercase\expandafter{\romannumeral1}_{2}}\otimes {\uppercase\expandafter{\romannumeral1}_{2}})\nonumber\\ &&\cdot \rho_{A B_{\uppercase\expandafter{\romannumeral1}} B_{\uppercase\expandafter{\romannumeral2}} C_{\uppercase\expandafter{\romannumeral1}} C_{\uppercase\expandafter{\romannumeral2}}}\nonumber\\ &&\cdot(E_{i} \otimes {\uppercase\expandafter{\romannumeral1}_{2}}\otimes {\uppercase\expandafter{\romannumeral1}_{2}}\otimes {\uppercase\expandafter{\romannumeral1}_{2}}\otimes {\uppercase\expandafter{\romannumeral1}_{2}})^{\dagger }.
\end{eqnarray}

Due to the disconnection between the interior region and the exterior region of the black hole, Alice, Bob, and Charlie cannot access the modes inside the event horizon. Therefore, the modes $B_{\uppercase\expandafter{\romannumeral1}}$ and $C_{\uppercase\expandafter{\romannumeral1}}$ outside the event horizon are called the accessible modes, and the modes $B_{\uppercase\expandafter{\romannumeral2}}$ and $C_{\uppercase\expandafter{\romannumeral2}}$ inside the event horizon are called the inaccessible modes. By tracing over the modes $B_{\uppercase\expandafter{\romannumeral2}}$ and $C_{\uppercase\expandafter{\romannumeral2}}$  on state ${\rho}''_{A B_{\uppercase\expandafter{\romannumeral1}} B_{\uppercase\expandafter{\romannumeral2}} C_{\uppercase\expandafter{\romannumeral1}} C_{\uppercase\expandafter{\romannumeral2}}}$ in Eq.(20), it is not difficult to get the physical accessible density matrix $\rho_{A B_{\uppercase\expandafter{\romannumeral1}} C_{\uppercase\expandafter{\romannumeral1}} }$ as 
\begin{equation}
\rho_{A B_{\uppercase\expandafter{\romannumeral1}} C_{\uppercase\expandafter{\romannumeral1}} }=\frac{1}{Z_{1}} \left( \begin{array}{cccccccc}
 \mu_{1} & 0 & 0 &0& 0 & 0& 0 & w_{1} \\
  0& \mu_{2} & 0 & 0& 0 & 0& 0 & 0\\
  0& 0 & \mu_{3} & 0& 0 & 0& 0 & 0\\
 0 & 0 &0  & \mu_{4}& 0 & 0& 0 & 0\\
 0 & 0 &0  &0& \nu_{4} & 0& 0 & 0\\
 0 & 0 &0  &0& 0 & \nu_{3}& 0 & 0\\
 0 & 0 &0  &0& 0 & 0& \nu_{2} & 0\\
 w^{*}_{1} & 0 &0  &0& 0 & 0& 0 & \nu_{1}
\end{array} \right),
\end{equation}
with
\begin{eqnarray}
&&\mu_{1}=-\frac{\alpha^{2} e^{\frac{2\omega_{k}}{T}} (-1+f) (1+e (-1+h))}{(1+e^{\frac{\omega_{k}}{T}})^{2}},\nonumber\\
&&\mu_{2}=\mu_{3}=-\frac{1}{4}\alpha^{2}(-1+f) (1+e (-1+h))  \mathrm{sech}\left [\frac{\omega_{k} }{2 T}\right ]^{2},\nonumber\\
&&\mu_{4}=(1-f)(e h+\alpha^{2}(\frac{1+e(-1+h)} {(1+e^{\frac{\omega_{k}}{T}})^{2} }-eh)),\nonumber\\
&&\nu_{1}=-\frac{\alpha^{2}e f(-1+h)} {(1+e^{\frac{\omega_{k}}{T}})^{2} }+(-1+\alpha^{2}) f (-1+e h),\nonumber\\
&&\nu_{2}=\nu_{3}=-\frac{1}{4}\alpha^{2} e f (-1+h) \mathrm{sech} \left [\frac{\omega_{k} }{2 T}\right ]^{2},\nonumber\\&&\nu_{4}=\frac{\alpha^{2}e f(1-h)} {(1+e^{-\frac{\omega_{k}}{T}})^{2}},\nonumber\\
&&w_{1}=\frac{\alpha \sqrt{1-\alpha^{2}} \sqrt{1-e} 
 e^{\frac{\omega_{k}}{T}}    \sqrt{-(-1+f)f}}{1+e^{\frac{\omega_{k}}{T}}},\nonumber\\
&&Z_{1}=f+(-1+2f) (\alpha^{2} (-1+e)-eh). 
\end{eqnarray}

According to Eqs.~(7) and (10), we can obtain the GTN and GTE of density matrix $\rho_{A B_{\uppercase\expandafter{\romannumeral1}} C_{\uppercase\expandafter{\romannumeral1}} }$. Due to the complicated expressions of these quantities, we do not attempt to write them out explicitly for simplicity.

In Fig.2, we plot the physically accessible GTN$(\rho_{A B_{\uppercase\expandafter{\romannumeral1}} C_{\uppercase\expandafter{\romannumeral1}} })$ and GTE$(\rho_{A B_{\uppercase\expandafter{\romannumeral1}} C_{\uppercase\expandafter{\romannumeral1}} })$ as functions of the Hawking temperature $T$ for different values of decoherence strength $r$. From the Fig.2(a), we can find that as the Hawking temperature $T$ increases, the GTN$(\rho_{A B_{\uppercase\expandafter{\romannumeral1}} C_{\uppercase\expandafter{\romannumeral1}} })$ will first be greater than 4 and then less than 4, which means that the ``sudden death'' of GTN$(\rho_{A B_{\uppercase\expandafter{\romannumeral1}} C_{\uppercase\expandafter{\romannumeral1}} })$ will take place, namely, the Hawking effect destroys the physically GTN which is initially shared by Alice, Bob and Charlie. By comparing the three lines in the figure, it is shown that with the increase of decoherence strength $r$, the initial value of GTN and the critical Hawking temperature $T_{c}$ for the sudden death of GTN will decrease, especially, when the decoherence strength $r$ is large enough, the value of GTN will always be less than 4. This phenomenon is not mentioned in the isolated system of Ref. \cite{a34a} and it has been observed first time that the decoherence completely destroys the physically accessible GTN which is shared by $A$, $B_{\uppercase\expandafter{\romannumeral1}} $ and $C_{\uppercase\expandafter{\romannumeral1}} $. Furthermore, we also find that the GTN$(\rho_{A B_{\uppercase\expandafter{\romannumeral1}} C_{\uppercase\expandafter{\romannumeral1}} })$ is not sensitive to the decoherence parameter $p$, that is, the decoherence parameter $p$ has little effect on the dynamics of GTN$(\rho_{A B_{\uppercase\expandafter{\romannumeral1}} C_{\uppercase\expandafter{\romannumeral1}} })$. From the Fig.2(b), it is found that the physically accessible GTE$(\rho_{A B_{\uppercase\expandafter{\romannumeral1}} C_{\uppercase\expandafter{\romannumeral1}} })$ shows a monotonic decrease as the Hawking temperature increases, and a stationary GTE  can survive as the Hawking temperature approaches infinity. This implies that the ``sudden death'' of GTE$(\rho_{A B_{\uppercase\expandafter{\romannumeral1}} C_{\uppercase\expandafter{\romannumeral1}} })$ will not occur in this situation for the infinite Hawking temperature, namely, the GTE$(\rho_{A B_{\uppercase\expandafter{\romannumeral1}} C_{\uppercase\expandafter{\romannumeral1}} })$ can be divided into two different parts: nonlocal and local. When the Hawking temperature $T$ exceeds the critical temperature $T_{c}$, the nonlocal GTE is destroyed completely by the Hawking effect, and the local GTE is preserved. Comparing the three lines in the figure, it is obvious that the initial value and the stationary value of GTE$(\rho_{A B_{\uppercase\expandafter{\romannumeral1}} C_{\uppercase\expandafter{\romannumeral1}} })$ will decline when decoherence strength $r$ increases. In particular, from the blue line of Fig.2(c), we can find that if the decoherence parameter $p$ is less than 1, the ``sudden death'' of GTE$(\rho_{A B_{\uppercase\expandafter{\romannumeral1}} C_{\uppercase\expandafter{\romannumeral1}} })$ will take place when the decoherence strength $r$ is large enough, which means that the local GTE is destroyed completely by the decoherence. The above results were not found in the isolated system studied in Ref. \cite{a34a} and suggest that in certain cases, the decoherence can completely destroy the physically accessible GTN and GTE, which is shared by Alice, Bob, and Charlie. This is very disadvantageous for the implementation of relativistic quantum information tasks.
\begin{figure}
\begin{center}
\includegraphics[width=4.0cm]{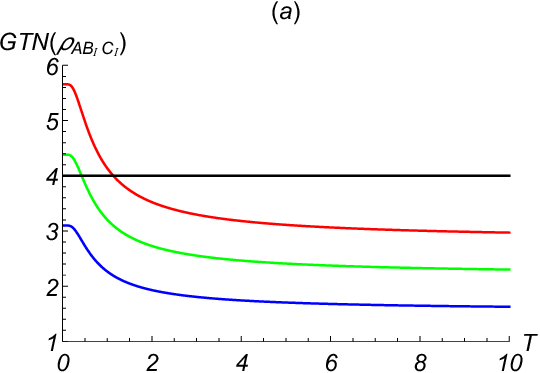}
\includegraphics[width=4.0cm]{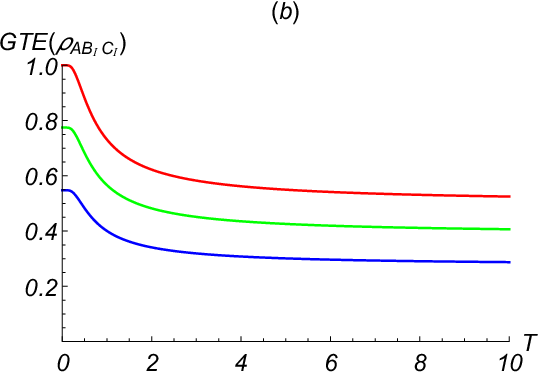}
\includegraphics[width=4.0cm]{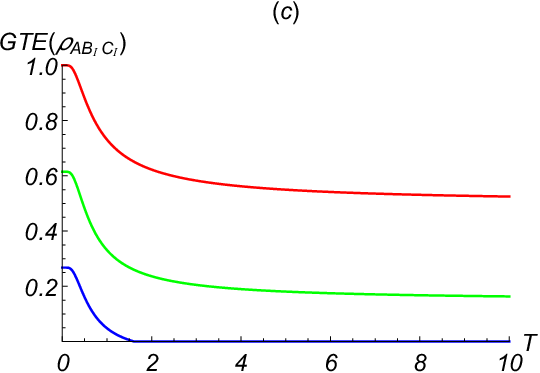}
\caption{\label{fig:fig1} (a) The physically accessible GTN$ (\rho _{A B_{\uppercase\expandafter{\romannumeral1}}C_{\uppercase\expandafter{\romannumeral1}}})$ as a function of the Hawking temperature $T$ with $\omega_{k}=1$,$\alpha=\frac{\sqrt{2}}{2}$,$ p=1$ and $f=0$ for different decoherence strength $r$: $r=0 $(red line), $r=0.4$(green  line), $r=0.7$(blue line). (b) The physically accessible GTE$ (\rho _{A B_{\uppercase\expandafter{\romannumeral1}}C_{\uppercase\expandafter{\romannumeral1}}})$ as a function of the Hawking temperature $T$ with $\omega_{k}=1$,$\alpha=\frac{\sqrt{2}}{2}$, $p=1$ and $f=0$ for different decoherence strength $r$: $r=0$ (red line), $r=0.4$(green line), $r=0.7$(blue line).(c)The physically accessible GTE$ (\rho _{A B_{\uppercase\expandafter{\romannumeral1}}C_{\uppercase\expandafter{\romannumeral1}}})$ as a function of the Hawking temperature $T$ with $\omega_{k}=1$,$\alpha=\frac{\sqrt{2}}{2}$ , $p=0.8$ and $f=0$ for different decoherence strength $r$: $r=0$ (red line), $r=0.4$(green line), $r=0.7$(blue line). }
\end{center}
\end{figure}

\begin{figure}
\begin{center}
\includegraphics[width=4.0cm]{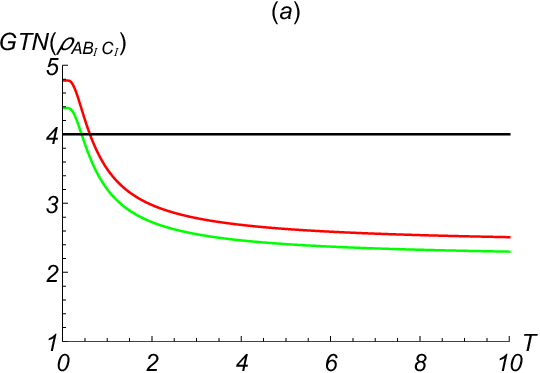}
\includegraphics[width=4.0cm]{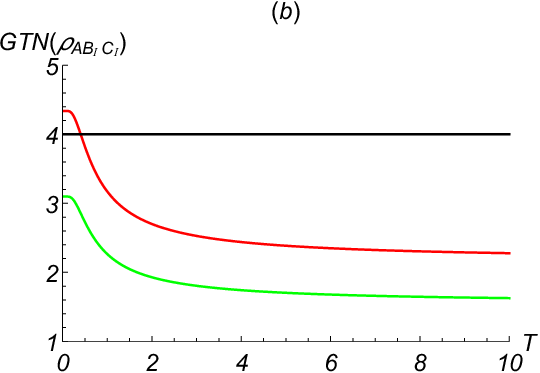}
\includegraphics[width=4.0cm]{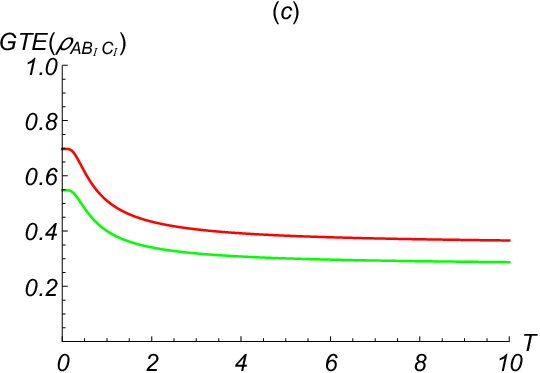}
\includegraphics[width=4.0cm]{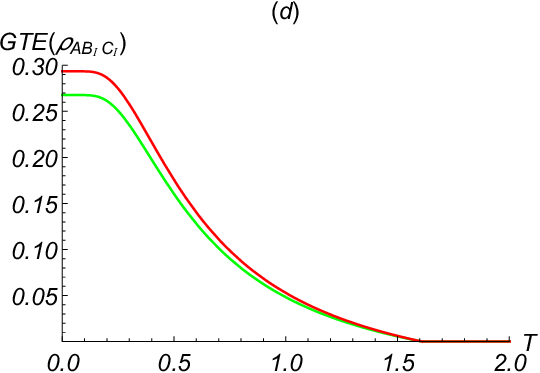}
\caption{\label{fig:fig1} (a) The physically accessible GTN$ (\rho _{A B_{\uppercase\expandafter{\romannumeral1}}C_{\uppercase\expandafter{\romannumeral1}}})$ as a function of the Hawking temperature $T$ with $r=0.4$, $\omega_{k}=1$,  $\alpha=\frac{\sqrt{2}}{2}$  and $p=1$ for different local filtering operation strength $f$: $f=0$ (green line), $f=0.7$(red line).(b) The physically accessible GTN$ (\rho _{A B_{\uppercase\expandafter{\romannumeral1}}C_{\uppercase\expandafter{\romannumeral1}}})$ as a function of the Hawking temperature $T$ with $r=0.7$,  $\omega_{k}=1$,  $\alpha=\frac{\sqrt{2}}{2}$  and $p=1$ for different local filtering operation strength $f$: $f=0$ (green line), $f=0.85$(red line).(c)The physically accessible GTE$ (\rho _{A B_{\uppercase\expandafter{\romannumeral1}}C_{\uppercase\expandafter{\romannumeral1}}})$ as a function of the Hawking temperature $T$  with  $r=0.7$, $\omega_{k}=1$,  $\alpha=\frac{\sqrt{2}}{2}$ and $p=1$  for different local filtering operation strength $f$:  $f=0$ (green line), $f=0.7$(red line).(d)The physically accessible GTE$ (\rho _{A B_{\uppercase\expandafter{\romannumeral1}}C_{\uppercase\expandafter{\romannumeral1}}})$ as a function of the Hawking temperature $T$ with  $r=0.7$, $\omega_{k}=1$,  $\alpha=\frac{\sqrt{2}}{2}$ and $p=0.8$ for different local filtering operation strength $f$: $f=0$ (green line), $f=0.75$(red line).}
\end{center}
\end{figure}

In order to explore the feasibility of a local filtering operation in suppressing the influence of the decoherence and Hawking effect, and effectively protecting the physically accessible GTN$(\rho_{A B_{\uppercase\expandafter{\romannumeral1}} C_{\uppercase\expandafter{\romannumeral1}} })$ and GTE$(\rho_{A B_{\uppercase\expandafter{\romannumeral1}} C_{\uppercase\expandafter{\romannumeral1}} })$.  In Fig.3, we plot the GTN$(\rho_{A B_{\uppercase\expandafter{\romannumeral1}} C_{\uppercase\expandafter{\romannumeral1}} })$ and GTE$(\rho_{A B_{\uppercase\expandafter{\romannumeral1}} C_{\uppercase\expandafter{\romannumeral1}} })$ as functions of the Hawking temperature $T$ with a local filtering operation. From the Fig.3(a), it is shown that when we perform a local filtering operation on Alice’s qubit, the value of physically accessible GTN$(\rho_{A B_{\uppercase\expandafter{\romannumeral1}} C_{\uppercase\expandafter{\romannumeral1}} })$ can be increased and the critical Hawking temperature of the ``sudden death'' of GTN$(\rho_{A B_{\uppercase\expandafter{\romannumeral1}} C_{\uppercase\expandafter{\romannumeral1}} })$ can be prolonged, that is, there is a larger range of Hawking temperature that the tripartite nonlocality GTN$(\rho_{A B_{\uppercase\expandafter{\romannumeral1}} C_{\uppercase\expandafter{\romannumeral1}} })$ which is shared by Alice, Bob and Charlie can exist by making use of the local filtering operation. In particular, comparing with the red line and the green line in Fig.3(b), we can find that even in the case of GTN$(\rho_{A B_{\uppercase\expandafter{\romannumeral1}} C_{\uppercase\expandafter{\romannumeral1}} })$ is completely destroyed by the decoherence and Hawking effect, we can make the tripartite nonlocality GTN$(\rho_{A B_{\uppercase\expandafter{\romannumeral1}} C_{\uppercase\expandafter{\romannumeral1}} })$ appear in a certain range of Hawking temperature by using the local filtering operation. From the green line of Fig.3(c), it is obvious that the ``sudden death'' of GTE$(\rho_{A B_{\uppercase\expandafter{\romannumeral1}} C_{\uppercase\expandafter{\romannumeral1}} })$ does not appear in this situation. Instead, the GTE$(\rho_{A B_{\uppercase\expandafter{\romannumeral1}} C_{\uppercase\expandafter{\romannumeral1}} })$ revives to a stable value after monotonic decreasing, which means that there is a stationary 
GTE$(\rho_{A B_{\uppercase\expandafter{\romannumeral1}} C_{\uppercase\expandafter{\romannumeral1}} })$ between Alice, Bob and Charlie as the Hawking temperature approaches infinity. Comparing the red line with the green line, we can see that the stable value of 
GTE$(\rho_{A B_{\uppercase\expandafter{\romannumeral1}} C_{\uppercase\expandafter{\romannumeral1}} })$ can be enhanced by applying the local filtering operation. In addition, comparing the red line with the green line of Fig.3(d), it is found that the value of
GTE$(\rho_{A B_{\uppercase\expandafter{\romannumeral1}} C_{\uppercase\expandafter{\romannumeral1}} })$ can be increased by performing the local filtering operation, but unfortunately the critical time of ``sudden death'' GTE$(\rho_{A B_{\uppercase\expandafter{\romannumeral1}} C_{\uppercase\expandafter{\romannumeral1}} })$ cannot be lengthened. These results have not been noted in previous studies \cite{a34a}, and indicate that the local filtering operation can be used to suppress the influence of decoherence and Hawking effect, and improve the physically accessible GTN$(\rho_{A B_{\uppercase\expandafter{\romannumeral1}} C_{\uppercase\expandafter{\romannumeral1}} })$ and GTE$(\rho_{A B_{\uppercase\expandafter{\romannumeral1}} C_{\uppercase\expandafter{\romannumeral1}} })$. This outcome may have potential applications in the implementation of relativistic quantum information tasks under decoherence.

For a better understanding of the behaviors of GTN and GTE for the tripartite subsystem in the system, we also discuss the physically inaccessible GTN and GTE of other tripartite subsystems. Tracing over the modes $C_{\uppercase\expandafter{\romannumeral1}}$ and $C_{\uppercase\expandafter{\romannumeral2}}$ on the state $ \rho^{''} _{A B_{\uppercase\expandafter{\romannumeral1}}B_{\uppercase\expandafter{\romannumeral2}}C_{\uppercase\expandafter{\romannumeral1}}C_{\uppercase\expandafter{\romannumeral2}}}$ in Eq.~(20), we obtain the reduced density matrix $ \rho _{A B_{\uppercase\expandafter{\romannumeral1}}B_{\uppercase\expandafter{\romannumeral2}}}$ as
\begin{equation}
\rho_{A B_{\uppercase\expandafter{\romannumeral1}} B_{\uppercase\expandafter{\romannumeral2}} }=\frac{1}{Z_{1}} \left( \begin{array}{cccccccc}
 \mu_{1} & 0 & 0 &0& 0 & 0& 0 & w_{1} \\
0& 0 & 0 & 0& 0 & 0& 0 & 0\\
0& 0 & \mu_{3} & 0& 0 & 0& 0 & 0\\
0 & 0 &0  & \mu_{4}& 0 & 0& 0 & 0\\
0 & 0 &0  &0& \nu_{4} & 0& 0 & 0\\
0 & 0 &0  &0& 0 & 0& 0 & 0\\
0 & 0 &0  &0& 0 & 0& \nu_{2} & 0\\
w^{\ast}_{1} & 0 &0  &0& 0 & 0& 0 & \nu_{1}
\end{array}\right),
\end{equation}
where
\begin{eqnarray*}
&\mu_{1}&=(1-f)(\frac{\alpha^{2}(1-e)(1-h)}{1+e^{-\frac{\omega_{k}}{T}}}+\frac{\alpha^{2} h}{1+e^{-\frac{\omega_{k}}{T}}}),\nonumber\\
&\mu_{3}&=(1-\alpha^{2})e(1-f)h,\nonumber\\
&\mu_{4}&=\frac{\alpha^{2}e(1-f)h}{1+e^{\frac{\omega_{k}}{T}}},\nu_{1}=f(\frac{\alpha^{2}(1-h)}{1+e^{\frac{\omega_{k}}{T}}}+\frac{\alpha^{2}(1-e)h}{1+e^{\frac{\omega_{k}}{T}}}),\nonumber\\
&\nu_{2}&=f((1-\alpha^{2})(1-h)+(1-\alpha^{2})(1-e)h),\nonumber\\
&\nu_{4}&=\frac{\alpha^{2}ef(1-h)}{1+e^{-\frac{\omega_{k}}{T}}},\nonumber\\
\end{eqnarray*}
\begin{eqnarray}
&w_{1}&=\sqrt{1-f}\sqrt{f}(\frac{\alpha^{2}\sqrt{1-e}(1-h)}{\sqrt{2+e^{-\frac{\omega_{k}}{T}}+e^{\frac{\omega_{k}}{T}}}}\nonumber\\
&+&\frac{\alpha^{2}\sqrt{1-e}h}{\sqrt{2+e^{-\frac{\omega_{k}}{T}}+e^{\frac{\omega_{k}}{T}}}}),\nonumber\\
&Z_{1}&=\frac{f+e h-2 e f h}{1+e^{\frac{\omega_{k}}{T}}}\nonumber\\&+&\frac{e^{\frac{\omega_{k}}{T}}(f+(-1+2f)(\alpha^{2}(-1+e)-eh))}{1+e^{\frac{\omega_{k}}{T}}}.
\end{eqnarray}

With the help of Eqs.~(7) and (10), we can obtain the GTN and GTE of density matrix $\rho_{A B_{\uppercase\expandafter{\romannumeral1}} B_{\uppercase\expandafter{\romannumeral2}} }$. Since the expressions for these quantities are more complex, we do not write them out explicitly for simplicity. Furthermore, according to the exchange symmetry for Bob and Charlie, we can get the GTN and GTE of density matrix $\rho_{A C_{\uppercase\expandafter{\romannumeral1}} C_{\uppercase\expandafter{\romannumeral2}}}$ are the same as those for the density matrix $\rho_{A B_{\uppercase\expandafter{\romannumeral1}} B_{\uppercase\expandafter{\romannumeral2}}}$. Thus, we just need to analyze the  GTN and GTE for density matrix $\rho_{A B_{\uppercase\expandafter{\romannumeral1}} B_{\uppercase\expandafter{\romannumeral2}}}$.
\begin{figure}
\begin{center}
\includegraphics[width=4.0cm]{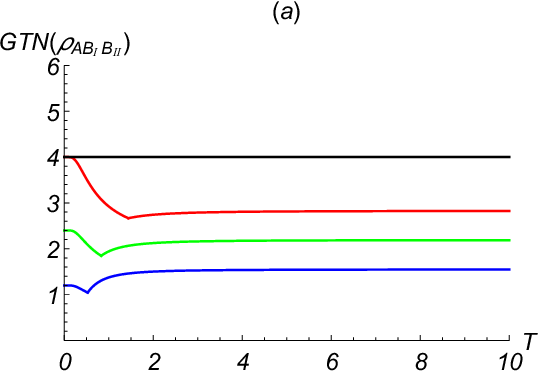}
\includegraphics[width=4.0cm]{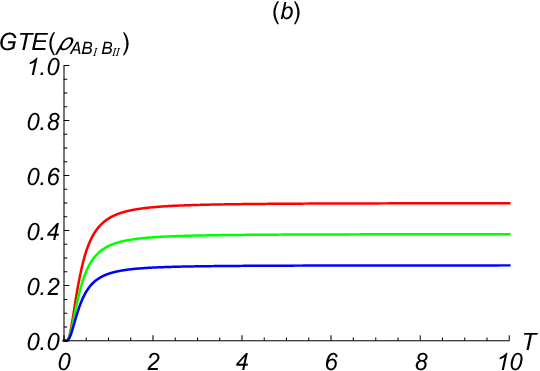}
\caption{\label{fig:fig1}  (a) The physically inaccessible GTN$ (\rho _{A B_{\uppercase\expandafter{\romannumeral1}}B_{\uppercase\expandafter{\romannumeral2}}})$ as a function of the Hawking temperature $T$ with $\omega_{k}=1$, $\alpha=\frac{\sqrt{2}}{2}$, $p=1$ and $f=0$ for different decoherence strength $r$: $r=0$ (red line), $r=0.4$(green line), $r=0.7$(blue line).(b) The physically inaccessible GTE$ (\rho _{A B_{\uppercase\expandafter{\romannumeral1}}B_{\uppercase\expandafter{\romannumeral2}}})$ as a function of the Hawking temperature $T$ with $\omega_{k}=1$, $\alpha=\frac{\sqrt{2}}{2}$, $p=1$ and $f=0$ for different decoherence strength $r$: $r=0 $(red line), $r=0.4$(green line), $r=0.7$(blue line).}
\end{center}
\end{figure}

In Fig.4, we plot GTN$(\rho_{A B_{\uppercase\expandafter{\romannumeral1}} B_{\uppercase\expandafter{\romannumeral2}}})$ and GTE$(\rho_{A B_{\uppercase\expandafter{\romannumeral1}} B_{\uppercase\expandafter{\romannumeral2}}})$  as functions of the Hawking temperature $T$ for different values of decoherence strength $r$ without local filtering operation. From the Fig.4(a), it is found that the GTN$(\rho_{A B_{\uppercase\expandafter{\romannumeral1}} B_{\uppercase\expandafter{\romannumeral2}}})$ is always less than 4 for any Hawking temperature $T$, which means that the physically inaccessible GTN between modes $A$, $B_{\uppercase\expandafter{\romannumeral1}}$, $B_{\uppercase\expandafter{\romannumeral2}}$ cannot be generated. In addition, it is obvious from the three lines of the figure that as the decoherence strength $r$ increases, the value of GTN will become smaller. From the Fig.4(b), we can see that the GTE$(\rho_{A B_{\uppercase\expandafter{\romannumeral1}} B_{\uppercase\expandafter{\romannumeral2}} })$ initially increases from zero and approaches to the stable value in the infinite Hawking temperature. This implies that the physically inaccessible GTE between modes $A$, $B_{\uppercase\expandafter{\romannumeral1}}$ and $B_{\uppercase\expandafter{\romannumeral2}}$ can be produced by the Hawking effect, even though they are separated by the event horizon of the black hole. The physical interpretation is that the $A$, $B_{\uppercase\expandafter{\romannumeral1}}$ and $B_{\uppercase\expandafter{\romannumeral2}}$ are entangled initially, and then the Hawking effect produce quantum entanglement between the modes $B_{\uppercase\expandafter{\romannumeral1}}$ and $B_{\uppercase\expandafter{\romannumeral2}}$, which is equivalent to an interaction between modes $B_{\uppercase\expandafter{\romannumeral1}}$ and $B_{\uppercase\expandafter{\romannumeral2}}$. This interaction induces some information to flow from mode $B_{\uppercase\expandafter{\romannumeral1}}$ to mode $B_{\uppercase\expandafter{\romannumeral2}}$. Therefore, the GTE of $\rho_{A B_{\uppercase\expandafter{\romannumeral1}} B_{\uppercase\expandafter{\romannumeral2}} }$ is established. Furthermore, as the strength of the Hawking effect increases, the physically inaccessible GTE increases rapidly. The different behaviors between GTN$(\rho_{A B_{\uppercase\expandafter{\romannumeral1}} B_{\uppercase\expandafter{\romannumeral2}}})$ and GTE$(\rho_{A B_{\uppercase\expandafter{\romannumeral1}} B_{\uppercase\expandafter{\romannumeral2}}})$ under the Hawking effect indicate that the nonlocality cannot pass through the event horizon of the black hole, while the entanglement can pass through the event horizon of black hole. Compare the three lines of the figure, it is shown that the appreciation of GTE$(\rho_{A B_{\uppercase\expandafter{\romannumeral1}} B_{\uppercase\expandafter{\romannumeral2}}})$ decelerates and the stable value decreases with the increase of decoherence strength $r$. This demonstrates that the decoherence may suppress the interaction between $B_{\uppercase\expandafter{\romannumeral1}}$ and $B_{\uppercase\expandafter{\romannumeral2}}$, decrease the information flow between $B_{\uppercase\expandafter{\romannumeral1}}$ and $B_{\uppercase\expandafter{\romannumeral2}}$, and reduce the generation of quantum entanglement between $B_{\uppercase\expandafter{\romannumeral1}}$ and $B_{\uppercase\expandafter{\romannumeral2}}$, that is, the decoherence will destroy the formation of the GTE between $A$,$B_{\uppercase\expandafter{\romannumeral1}}$ and $B_{\uppercase\expandafter{\romannumeral2}}$.

\begin{figure}
\begin{center}
\includegraphics[width=4.0cm]{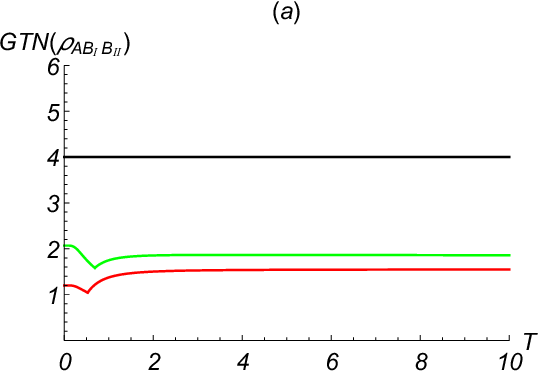}
\includegraphics[width=4.0cm]{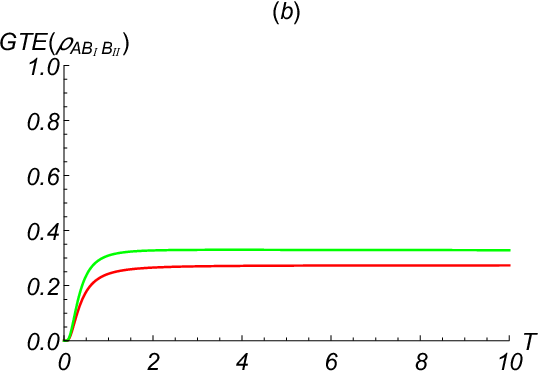}
\caption{\label{fig:fig1}  (a) The physically inaccessible GTN$ (\rho _{A B_{\uppercase\expandafter{\romannumeral1}}B_{\uppercase\expandafter{\romannumeral2}}})$ as a function of the Hawking temperature $T$ with $r=0.7$, $\omega_{k}=1$, $\alpha=\frac{\sqrt{2}}{2}$ and $p=1$ for different local filtering operation strength $f$: $f=0$(red line), $f=0.8$(green line).(b) The physically inaccessible GTE$ (\rho _{A B_{\uppercase\expandafter{\romannumeral1}}B_{\uppercase\expandafter{\romannumeral2}}})$ as a function of the Hawking temperature $T$ with $r=0.7$, $\omega_{k}=1$, $\alpha=\frac{\sqrt{2}}{2}$ and $p=1$ for different local filtering operation strength $f$: $f=0$ (red line), $f=0.8$(green line).}
\end{center}
\end{figure}

In Fig.5, we display the influence of local filtering operation on the dynamics of GTN and GTE of density matrix  $\rho_{A B_{\uppercase\expandafter{\romannumeral1}} B_{\uppercase\expandafter{\romannumeral2}}}$ under decoherence. From the Fig.5(a), it is found that the value of GTN$(\rho_{A B_{\uppercase\expandafter{\romannumeral1}} B_{\uppercase\expandafter{\romannumeral2}}})$ can be increased by local filtering operation, but it cannot exceed 4 for for any Hawking temperature $T$, namely, the GTN$(\rho_{A B_{\uppercase\expandafter{\romannumeral1}} B_{\uppercase\expandafter{\romannumeral2}}})$ cannot be produced, even though the local filtering operation is performed. From the red line and green line of Fig.5(b), we can see that the appreciation of GTE$(\rho_{A B_{\uppercase\expandafter{\romannumeral1}} B_{\uppercase\expandafter{\romannumeral2}}})$ accelerates and the stable value increases when the local filtering operation is performed. This means that the local filtering operation can suppress the influence of decoherence, slow down the exchange of the quantum information between the subsystem $A B_{\uppercase\expandafter{\romannumeral1}} B_{\uppercase\expandafter{\romannumeral2}}$ and environments, and accelerate the generation of quantum entanglement between $B_{\uppercase\expandafter{\romannumeral1}}$ and $B_{\uppercase\expandafter{\romannumeral2}}$. Therefore, the GTE between the modes $A$, $B_{\uppercase\expandafter{\romannumeral1}}$ and $B_{\uppercase\expandafter{\romannumeral2}}$ is enhanced. 

Next, we discuss the physically inaccessible GTN and GTE between the modes $A$, $B_{\uppercase\expandafter{\romannumeral2}}$ and $C_{\uppercase\expandafter{\romannumeral2}}$. By tracing over the modes $B_{\uppercase\expandafter{\romannumeral1}}$ and $C_{\uppercase\expandafter{\romannumeral1}}$ on the state $\rho^{''}_{A B_{\uppercase\expandafter{\romannumeral1}} B_{\uppercase\expandafter{\romannumeral2}}
C_{\uppercase\expandafter{\romannumeral1}} C_{\uppercase\expandafter{\romannumeral2}}}$ in Eq.~(20), the reduced density matrix 
$\rho_{A B_{\uppercase\expandafter{\romannumeral2}}
C_{\uppercase\expandafter{\romannumeral2}}}$can be expressed as

\begin{equation}
\rho_{A B_{\uppercase\expandafter{\romannumeral2}} C_{\uppercase\expandafter{\romannumeral2}} }=\frac{1}{Z_{1}} \left( \begin{array}{cccccccc}
 \mu_{1} & 0 & 0 &0& 0 & 0& 0 & 0 \\
  0& \mu_{2}& 0 & 0& 0 & 0& 0 & 0\\
  0& 0 & \mu_{3} & 0& 0 & 0& 0 & 0\\
 0 & 0 &0  & \mu_{4}& w_{4} & 0& 0 & 0\\
 0 & 0 &0  &w^{\ast}_{4}& \nu_{4} & 0& 0 & 0\\
 0 & 0 &0  &0& 0 & \nu_{3}& 0 & 0\\
 0 & 0 &0  &0& 0 & 0& \nu_{2} & 0\\
 0 & 0 &0  &0& 0 & 0& 0 & \nu_{1}
\end{array} \right),
\end{equation}
with
\begin{eqnarray}
&\mu_{1}&=(1-f)(\frac{\alpha^{2}(1-e)(1-h)}{(1+e^{-\frac{\omega_{k}}{T}})^{2}}+(1-\alpha^{2})eh\nonumber\nonumber\\&+&\frac{\alpha^{2} h}{(1+e^{-\frac{\omega_{k}}{T}})^{2}}),\nonumber\\
&\mu_{2}&=\mu_{3}=(1-f)(\frac{\alpha^{2}(1-e)(1-h)}{2+e^{-\frac{\omega_{k}}{T}}+e^{\frac{\omega_{k}}{T}}}+
\frac{\alpha^{2}h}{2+e^{-\frac{\omega_{k}}{T}}+e^{\frac{\omega_{k}}{T}}}),\nonumber\\
&\mu_{4}&=(1-f) (\frac{\alpha^{2}(1-e)(1-h)}{(1+e^{\frac{\omega_{k}}{T}})^{2}}+
\frac{\alpha^{2}h}{(1+e^{\frac{\omega_{k}}{T}})^{2}}),\nonumber\\
&\nu_{1}&= 
\frac{\alpha^{2}ef(1-h)}{(1+e^{\frac{\omega_{k}}{T}})^{2}},\nonumber\\
&\nu_{2}&=\nu_{3}=
\frac{\alpha^{2}ef(1-h)}{2+e^{-\frac{\omega_{k}}{T}}+e^{\frac{\omega_{k}}{T}}},\nonumber\\
&\nu_{4}&= f((1-\alpha^{2})(1-h)+\frac{\alpha^{2}e(1-h)}{(1+e^{-\frac{\omega_{k}}{T}})^{2}}+\nonumber\\&(&1-\alpha^{2})(1-e)h),\nonumber\\
&w_{4}&=\sqrt{1-f}\sqrt{f}(\frac{\alpha\sqrt{1-\alpha^{2}}\sqrt{1-e}(1-h)}{1+e^{\frac{\omega_{k}}{T}}}\nonumber\\&+&
\frac{\alpha\sqrt{1-\alpha^{2}}\sqrt{1-e}h}{1+e^{\frac{\omega_{k}}{T}}}),\nonumber\\
&Z_{1}&=f+(-1+2f)(\alpha^{2}(-1+e)-eh).    
\end{eqnarray}

Substituting the Eq.~(26) into the Eqs.~(7) and (10), the GTN and GTE of density matrix $\rho_{A B_{\uppercase\expandafter{\romannumeral2}}
C_{\uppercase\expandafter{\romannumeral2}}}$ can be obtained. For simplicity, we do not report the expressions of them.

\begin{figure}
\begin{center}
\includegraphics[width=4.0cm]{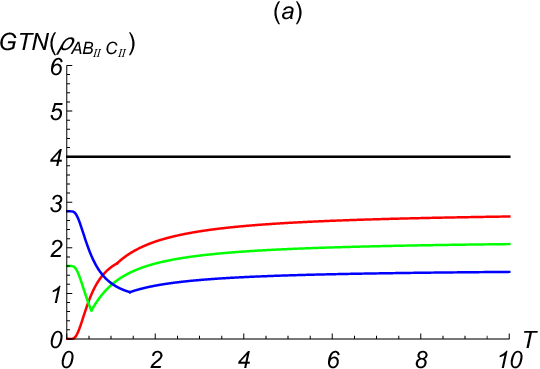}
\includegraphics[width=4.0cm]{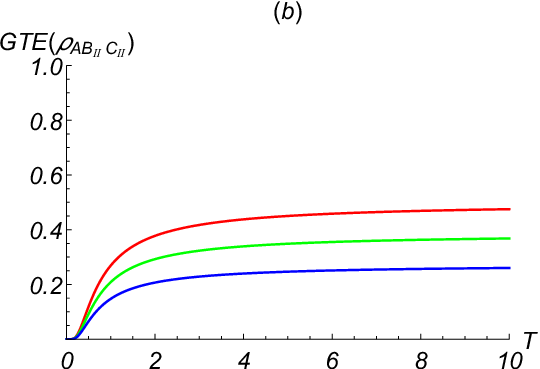}
\caption{\label{fig:fig1}  (a) The physically inaccessible GTN$ (\rho _{A B_{\uppercase\expandafter{\romannumeral2}}C_{\uppercase\expandafter{\romannumeral2}}})$ as a function of the Hawking temperature $T$ with $\omega_{k}=1$, $\alpha=\frac{\sqrt{2}}{2}$, $p=1$ and $f=0$ for different decoherence strength $r$: $r=0$ (red line), $r=0.4$(green  line), $r=0.7$(blue line).(b) The physically inaccessible GTE$ (\rho _{A B_{\uppercase\expandafter{\romannumeral2}}C_{\uppercase\expandafter{\romannumeral2}}})$ as a function of the Hawking temperature $T$ with $\omega_{k}=1$, $\alpha=\frac{\sqrt{2}}{2}$, $p=1$ and $f=0$ for different decoherence strength $r$: $r=0$ (red  line), $r=0.4$(green line), $r=0.7$(blue line).}
\end{center}
\end{figure}

In Fig.6, the GTN$(\rho_{A B_{\uppercase\expandafter{\romannumeral2}} C_{\uppercase\expandafter{\romannumeral2}}})$ and GTE$(\rho_{A B_{\uppercase\expandafter{\romannumeral2}} C_{\uppercase\expandafter{\romannumeral2}}})$ are displayed as functions of the Hawking temperature $T$ for different values of decoherence strength $r$ without local filtering operation. We can see from the Fig.6(a) that although the value of GTN$(\rho_{A B_{\uppercase\expandafter{\romannumeral2}} C_{\uppercase\expandafter{\romannumeral2}}})$ is not zero, it will always be less than 4 for any Hawking temperature $T$, which demonstrates that the physically inaccessible GTN between modes $A$, $B_{\uppercase\expandafter{\romannumeral2}}$ and $C_{\uppercase\expandafter{\romannumeral2}}$ cannot be generated, that is, the nonlocality cannot pass through the event horizon of black hole. By comparing the three lines in the figure, it is shown that the value of GTN$(\rho_{A B_{\uppercase\expandafter{\romannumeral2}} C_{\uppercase\expandafter{\romannumeral2}}})$ decreases with increase of decoherence strength $r$, which means that the generation of nonlocality will be disrupted by decoherence. In addition, we can find from the Fig6(b) that the physically inaccessible GTE between modes $A$, $B_{\uppercase\expandafter{\romannumeral2}}$ and $C_{\uppercase\expandafter{\romannumeral2}}$ can be generated, and a stationary GTE$(\rho_{A B_{\uppercase\expandafter{\romannumeral2}} C_{\uppercase\expandafter{\romannumeral2}}})$ can appear as the Hawking temperature approaches infinity. This implies that the entanglement can pass through the event horizon of the black hole, namely, the initial quantum entanglement in the physically accessible region flows into the physically inaccessible region. Compare the three lines of the figure, it is obvious that the increase rate and stable value of GTE$(\rho_{A B_{\uppercase\expandafter{\romannumeral2}} C_{\uppercase\expandafter{\romannumeral2}}})$ will decrease as the decoherence increases, which means that the flow of quantum entanglement from the physically accessible region to the physically inaccessible region will be destroyed by decoherence. It is worth noting that when the black hole evaporates completely, the asymptotic value of
GTE$(\rho_{A B_{\uppercase\expandafter{\romannumeral1}} C_{\uppercase\expandafter{\romannumeral1}}})$ and GTE$(\rho_{A B_{\uppercase\expandafter{\romannumeral2}} C_{\uppercase\expandafter{\romannumeral2}}})$ are equal for any decoherence strength $r$.
\begin{figure}
\begin{center}
\includegraphics[width=4.0cm]{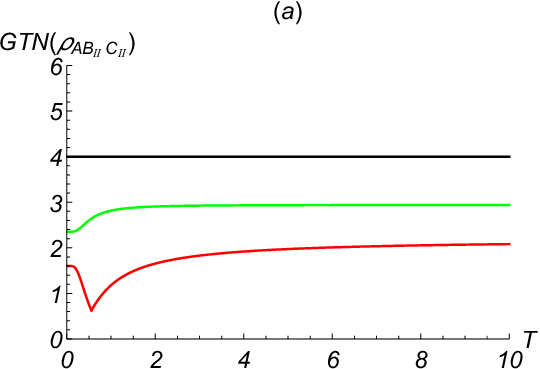}
\includegraphics[width=4.0cm]{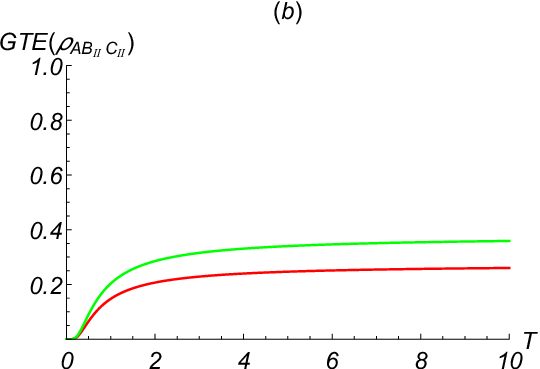}
\caption{\label{fig:fig1} (a) The physically inaccessible GTN$ (\rho _{A B_{\uppercase\expandafter{\romannumeral2}}C_{\uppercase\expandafter{\romannumeral2}}})$ as a function of the Hawking temperature $T$ with $r=0.4$, $\omega_{k}=1$, $\alpha=\frac{\sqrt{2}}{2}$ and  $p=1$ for different local filtering operation strength $f$: $f=0$ (red line), $f=0.9$(green line).(b) The physically inaccessible GTE$ (\rho _{A B_{\uppercase\expandafter{\romannumeral2}}C_{\uppercase\expandafter{\romannumeral2}}})$ as a function of the Hawking temperature $T$ with $r=0.7$, $\omega_{k}=1$, $\alpha=\frac{\sqrt{2}}{2}$  and $p=1$ for different local filtering operation strength $f$: $f=0$ (red line), $f=0.8$(green line).}
\end{center}
\end{figure}

In Fig.7, we plot the GTN$(\rho_{A B_{\uppercase\expandafter{\romannumeral2}} C_{\uppercase\expandafter{\romannumeral2}}})$ and GTE$(\rho_{A B_{\uppercase\expandafter{\romannumeral2}} C_{\uppercase\expandafter{\romannumeral2}}})$ as functions of the Hawking temperature $T$ with local filtering operation under decoherence. From the Fig.7, we can find similar results which are observed in the tripartite subsystem $A B_{\uppercase\expandafter{\romannumeral1}} B_{\uppercase\expandafter{\romannumeral2}}$. It is seen from the Fig.7(a) that the GTN$(\rho_{A B_{\uppercase\expandafter{\romannumeral2}} C_{\uppercase\expandafter{\romannumeral2}}})$ is less than 4 for any Hawking temperature $T$, even if we apply the local filtering operation, which means that the nonlocality cannot pass through the event horizon of black hole, even though the local filtering operation is performed. Furthermore, we can also find that the growth rate and the stable value of GTE$(\rho_{A B_{\uppercase\expandafter{\romannumeral2}} C_{\uppercase\expandafter{\romannumeral2}}})$ can be increased by applying the local filtering operation for different decoherence strength $r$. This indicates that the flow of quantum entanglement from the physically accessible region to the physically inaccessible region can be enhanced by using the local filtering operation. 

Finally, we explore the physically inaccessible GTN and GTE for the subsystem $A B_{\uppercase\expandafter{\romannumeral1}} C_{\uppercase\expandafter{\romannumeral2}}$. Tracing over the modes $B_{\uppercase\expandafter{\romannumeral2}}$ and $C_{\uppercase\expandafter{\romannumeral1}}$ on the state $\rho^{''}_{A B_{\uppercase\expandafter{\romannumeral1}} B_{\uppercase\expandafter{\romannumeral2}}
C_{\uppercase\expandafter{\romannumeral1}} C_{\uppercase\expandafter{\romannumeral2}}}$ in Eq.~(20), the expression of reduced density matrix $\rho_{A B_{\uppercase\expandafter{\romannumeral1}} C_{\uppercase\expandafter{\romannumeral2}}}$ can be written as

\begin{equation}
\rho_{A B_{\uppercase\expandafter{\romannumeral1}} C_{\uppercase\expandafter{\romannumeral2}} }=\frac{1}{Z_{1}} \left( \begin{array}{cccccccc}
 \mu_{1} & 0 & 0 &0& 0 & 0& 0 & 0 \\
  0& \mu_{2}& 0 & 0& 0 & 0& w_{2} & 0\\
  0& 0 & \mu_{3} & 0& 0 & 0& 0 & 0\\
 0 & 0 &0  & \mu_{4}& 0 & 0& 0 & 0\\
 0 & 0 &0  &0& \nu_{4} & 0& 0 & 0\\
 0 & 0 &0  &0& 0 & \nu_{3}& 0 & 0\\
 0 & w^{\ast}_{2} &0  &0& 0 & 0& \nu_{2} & 0\\
 0 & 0 &0  &0& 0 & 0& 0 & \nu_{1}
\end{array} \right),
\end{equation}
with
\begin{eqnarray}
&\mu_{1}&=(1-f)(\frac{\alpha^{2}(1-e)(1-h)}{(1+e^{-\frac{\omega_{k}}{T}})^{2}}+\frac{\alpha^{2} h}{(1+e^{-\frac{\omega_{k}}{T}})^{2}}),\nonumber\\
&\mu_{2}&=(1-f)(\frac{\alpha^{2}(1-e)(1-h)}{2+e^{-\frac{\omega_{k}}{T}}+e^{\frac{\omega_{k}}{T}}}+
\frac{\alpha^{2}h}{2+e^{-\frac{\omega_{k}}{T}}+e^{\frac{\omega_{k}}{T}}}),\nonumber\\
&\mu_{3}&=(1-f)(\frac{\alpha^{2}(1-e)(1-h)}{2+e^{-\frac{\omega_{k}}{T}}+e^{\frac{\omega_{k}}{T}}}+(1-\alpha^{2})eh\nonumber\\&+&
\frac{\alpha^{2}h}{2+e^{-\frac{\omega_{k}}{T}}+e^{\frac{\omega_{k}}{T}}}),\nonumber\\
&\mu_{4}&=(1-f)(\frac{\alpha^{2}(1-e)(1-h)}{(1+e^{\frac{\omega_{k}}{T}})^{2}}+\frac{\alpha^{2} h}{(1+e^{\frac{\omega_{k}}{T}})^{2}}),
\nonumber\\&\nu_{1}&=\frac{\alpha^{2} e f (1-h)}{(1+e^{\frac{\omega_{k}}{T}})^{2}},
\nonumber\\&\nu_{2}&=f ((1-\alpha^{2})(1-h)+\frac{\alpha^{2}e (1-h)}{2+e^{-\frac{\omega_{k}}{T}}+e^{\frac{\omega_{k}}{T}}}\nonumber\\
&+&(1-\alpha^{2})(1-e)h),\nonumber\\
&\nu_{3}&=\frac{\alpha^{2}e f (1-h)}{2+e^{-\frac{\omega_{k}}{T}}+e^{\frac{\omega_{k}}{T}}}, \nonumber\\
&\nu_{4}&=\frac{\alpha^{2} e f (1-h)}{(1+e^{-\frac{\omega_{k}}{T}})^{2}}, \nonumber\\
&w_{2}&=\sqrt{1-f}\sqrt{f}(\frac{\alpha \sqrt{1-\alpha^{2}} \sqrt{1-e}(1-h)}{\sqrt{2+e^{-\frac{\omega_{k}}{T}}+e^{\frac{\omega_{k}}{T}}}} \nonumber\\
&+&\frac{\alpha \sqrt{1-\alpha^{2}} \sqrt{1-e}h}{\sqrt{2+e^{-\frac{\omega_{k}}{T}}+e^{\frac{\omega_{k}}{T}}}}), \nonumber\\
&Z_{1}&=f+(-1+2f)(\alpha^{2}(-1+e)-eh).    
\end{eqnarray}

Inserting the Eq.~(28) into the Eqs. (7) and (10), the GTN and GTE of density matrix $\rho_{A B_{\uppercase\expandafter{\romannumeral1}} C_{\uppercase\expandafter{\romannumeral2}}}$ can be obtained. Here, we do not display the accurate expressions of GTN and GTE, since they are more complex. Furthermore, according to the exchange symmetry for Bob and Charlie, we can derive the GTN and GTE of density matrix $\rho_{A B_{\uppercase\expandafter{\romannumeral2}} C_{\uppercase\expandafter{\romannumeral1}}}$ are the same as those for the density matrix $\rho_{A B_{\uppercase\expandafter{\romannumeral1}} C_{\uppercase\expandafter{\romannumeral2}}}$. Thus, we just need to analyze the GTN and GTE for density matrix $\rho_{A B_{\uppercase\expandafter{\romannumeral1}} C_{\uppercase\expandafter{\romannumeral2}}}$.

\begin{figure}
\begin{center}
\includegraphics[width=4.0cm]{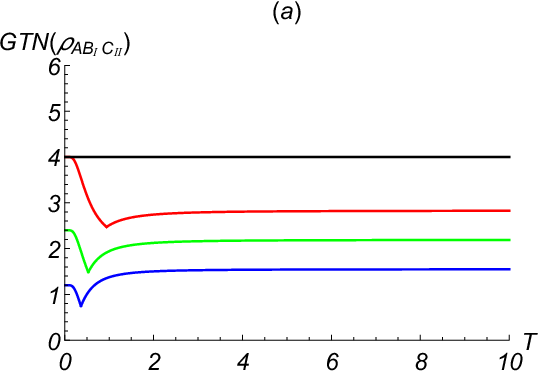}
\includegraphics[width=4.0cm]{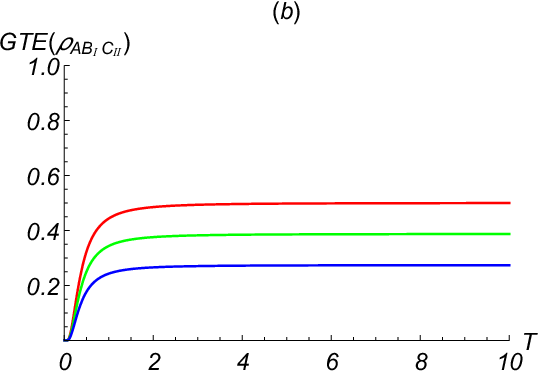}
\caption{\label{fig:fig1}   (a) The physically inaccessible GTN$ (\rho _{A B_{\uppercase\expandafter{\romannumeral1}}C_{\uppercase\expandafter{\romannumeral2}}})$  as a function of the Hawking temperature $T$ with $\omega_{k}=1$, $\alpha=\frac{\sqrt{2}}{2}$, $p=1$ and $f=0$ for different decoherence strength $r$: $r=0 $(red line), $r=0.4$(green  line), $r=0.7$(blue line).(b) The physically inaccessible GTE$ (\rho _{A B_{\uppercase\expandafter{\romannumeral1}}C_{\uppercase\expandafter{\romannumeral2}}})$ as a function of the Hawking temperature $T$ with $\omega_{k}=1$, $\alpha=\frac{\sqrt{2}}{2}$, $p=1$ and $f=0$ for different decoherence strength $r$: $r=0$ (red line), $r=0.4$(green line), $r=0.7$(blue line).}
\end{center}
\end{figure}

In Fig.8, we plot the GTN$(\rho_{A B_{\uppercase\expandafter{\romannumeral1}} C_{\uppercase\expandafter{\romannumeral2}}})$ and GTE$(\rho_{A B_{\uppercase\expandafter{\romannumeral1}} C_{\uppercase\expandafter{\romannumeral2}}})$ between Alice, Bob and anti-Charlie as functions of the Hawking temperature $T$ for different decoherence strength $r$ without local filtering operation. It is shown that with the increase of Hawking temperature $T$, the GTN$(\rho_{A B_{\uppercase\expandafter{\romannumeral1}} C_{\uppercase\expandafter{\romannumeral2}}})$ has always been less 4, and the GTE$(\rho_{A B_{\uppercase\expandafter{\romannumeral1}} C_{\uppercase\expandafter{\romannumeral2}}})$ increases from the zero and reaches a stable value for $T\to \infty $ , which means that the physically inaccessible GTE between modes $A$ , 
 $B_{\uppercase\expandafter{\romannumeral1}} $   and $C_{\uppercase\expandafter{\romannumeral2}}$ can be produced, but the physically inaccessible GTN between them cannot be produced. These indicate that the entanglement can pass through the event horizon of the black hole, but the nonlocality cannot. Furthermore, from the three lines of Fig.8(b), we can see that the increasing degree and the stable value of GTE$(\rho_{A B_{\uppercase\expandafter{\romannumeral1}} C_{\uppercase\expandafter{\romannumeral2}}})$  will decrease as the decoherence strength $r$ increases, which implies that the generation of the physically inaccessible GTE$(\rho_{A B_{\uppercase\expandafter{\romannumeral1}} C_{\uppercase\expandafter{\romannumeral2}}})$ will be destroyed by decoherence.  It's also worth pointing out that when the black hole evaporates completely, the asymptotic value of GTE$(\rho_{A B_{\uppercase\expandafter{\romannumeral1}} C_{\uppercase\expandafter{\romannumeral2}}})$, GTE$(\rho_{A B_{\uppercase\expandafter{\romannumeral1}} C_{\uppercase\expandafter{\romannumeral1}}})$,
GTE$(\rho_{A B_{\uppercase\expandafter{\romannumeral1}} C_{\uppercase\expandafter{\romannumeral2}}})$  are equal for any decoherence strength $r$.

\begin{figure}
\begin{center}
\includegraphics[width=4.0cm]{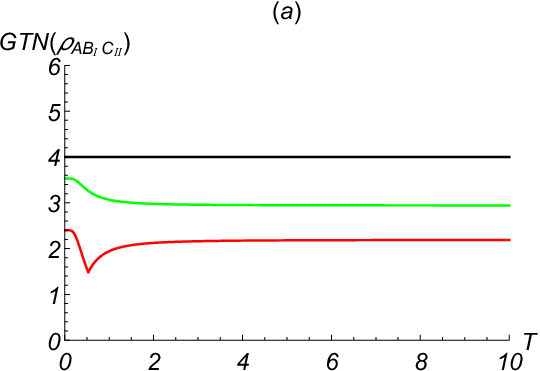}
\includegraphics[width=4.0cm]{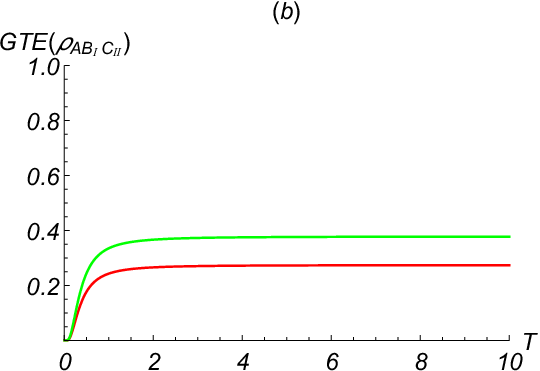}
\caption{\label{fig:fig1}  (a) The physically inaccessible GTN$ (\rho _{A B_{\uppercase\expandafter{\romannumeral1}}C_{\uppercase\expandafter{\romannumeral2}}})$ as a function of the Hawking temperature $T$ with $r=0.4$, $\omega_{k}=1$, $\alpha=\frac{\sqrt{2}}{2}$ and $p=1$ for different local filtering operation strength $f$: $f=0$ (red line), $f=0.9$(green line).(b) The physically inaccessible GTE$ (\rho _{A B_{\uppercase\expandafter{\romannumeral1}}C_{\uppercase\expandafter{\romannumeral2}}})$ as a function of the Hawking temperature $T$ with $r=0.7$, $\omega_{k}=1$, $\alpha=\frac{\sqrt{2}}{2}$  and $p=1$ for different local filtering operation strength $f$: $f=0$ (red line), $f=0.8$(green line).}
\end{center}
\end{figure}

In Fig.9, we demonstrate the effect of the local filtering operation on the dynamics of GTN and GTE for the density matrix $\rho_{A B_{\uppercase\expandafter{\romannumeral1}} C_{\uppercase\expandafter{\romannumeral2}}}$ under decoherence. It can be seen from the Fig.9(b) that the value of GTN$(\rho_{A B_{\uppercase\expandafter{\romannumeral1}} C_{\uppercase\expandafter{\romannumeral2}}})$ cannot exceed 4 for any Hawking temperature $T$, that is, the GTN$(\rho_{A B_{\uppercase\expandafter{\romannumeral1}} C_{\uppercase\expandafter{\romannumeral2}}})$ cannot be produced, even if we perform the local filtering operation. In addition, from the Fig.9(b), it is shown that when we apply the local filtering operation, the growth rate of GTE$(\rho_{A B_{\uppercase\expandafter{\romannumeral1}} C_{\uppercase\expandafter{\romannumeral2}}})$ will be accelerated, and the stable value of GTE$(\rho_{A B_{\uppercase\expandafter{\romannumeral1}} C_{\uppercase\expandafter{\romannumeral2}}})$  can be enhanced, even under decoherence. This result implies that the local filtering operation can suppress the exchange of the quantum information between the subsystem $A  
 B_{\uppercase\expandafter{\romannumeral1}} C_{\uppercase\expandafter{\romannumeral2}}$ and environments, and accelerate the flow of quantum entanglement from the physically accessible region to the physically inaccessible region.

\section{CONCLUSIONS}
In this paper, we discuss the impact of the environment on the GTN and GTE of the system in the background of the Schwarzschild black
hole and explore the amplification of the GTN and GTE of Dirac particles under decoherence by a local filtering operation. It is demonstrated the “sudden death” of the physically accessible GTN takes place at some critical Hawking temperature, and the critical Hawking temperature degrades with the rise of the decoherence strength. In particular, we find that the physically accessible GTN will be destroyed by decoherence, which means that the physically accessible GTN will not exist in the system. It is worth noting that the local filtering operation can generate the physically accessible GTN within a certain range of Hawking temperature, even if the GTN of the system is destroyed by decoherence. This result has been observed first time and is beneficial for the realization of GTN-based quantum information tasks.

Furthermore, we also see that the physically accessible GTE approaches the nonzero asymptotic value in the limit of infinite Hawking temperature, which means that the “sudden death” of the physically accessible GTE does not exist in the system, when the impact of the environment is not too large. However, we first time show that if the decoherence parameter p is less than 1, the “sudden death” of physically accessible GTE will take place when the decoherence strength is large enough, which is different from the previous research. It is worth noting that the decay speed of the physically accessible GTE decreases, and the stable value of the physically accessible GTE improves if we perform the local filtering operation. These results may have potential applications in the implementation of relativistic quantum information tasks under decoherence.

Finally, we discuss the production of physically inaccessible GTN and GTE of other tripartite subsystems. It is shown that with the rise of the Hawking temperature, the physically accessible GTN decreases, but the physically inaccessible GTN is not generated. However, the behaviors of physically inaccessible GTE are different. With the loss of the physically accessible GTE, the physically inaccessible GTE initially increases from zero and approaches the stable value in the infinite Hawking temperature, which means that the physically inaccessible GTE can be produced by the Hawking effect. Meanwhile, it can also be seen that when the black hole evaporates completely, the stable value of physically accessible GTE and physically inaccessible GTE are equal for any decoherence strength. These results imply that the GTE is redistributive, but GTN cannot be redistributed through the Hawking effect, namely, the quantum entanglement can pass through the event horizon of the black hole, but the quantum nonlocality can not. It is worth noting that the growth rate and stable value of physically inaccessible GTE decrease as the decoherence increases, which demonstrates that the flow of quantum entanglement from the physically accessible region to the physically inaccessible region would be suppressed, and the redistribution of GTE would be destroyed by the decoherence. We also find that the growth rate of the physically inaccessible GTE and the stable value of the physically inaccessible GTE can be increased by performing the local filtering operation. The results of this paper may enrich the study of tripartite quantum properties from isolated systems to open systems under the framework of relativity. 

\begin{acknowledgments}
This project was supported by the National Natural Science
Foundation of China (Grant Nos. 12265007 and 11364006).
\end{acknowledgments}


\begin{thebibliography}{99}
\bibitem{a1a} A. Einstein, B. Podolsky, and N. Rosen, Can Quantum-Mechanical Description of Physical Reality Be Considered Complete?, Phys. Rev. \textbf{47}, 777 (1935).
\bibitem{a2a} J. S. Bell, On the Einstein-Podolsky-Rosen paradox, Physics \textbf{1}, 195 (1964).
\bibitem{a3a} N. Brunner,  D. Cavalcanti, S. Pironio,  V. Scarani, and S. Wehner, Bell nonlocality, Rev. Mod. Phys. \textbf{86}, 419 (2014).
\bibitem{a4a} A. Aspect, P. Grangier, and G. Roger, Experimental tests of realistic local theories via Bell’s theorem, Phys. Rev. Lett. \textbf{47}, 460 (1981).
\bibitem{a5a} B. Hensen, H. Bernien, A. E. Dr\'{e}au,  et al., Loophole-free Bell inequality violation using electron spins separated by 1.3 km,
Nature \textbf{526}, 682 (2015).
\bibitem{a6a}  J. Handsteiner,  A. S. Friedman, D. Rauch, et al., Cosmic Bell test measurement settings from Milky Way stars, Phys. Rev. Lett.
\textbf{118}, 060401 (2017).
\bibitem{a7a} H. Buhrman, R. Cleve, S. Massar, and R.D. Wolf, Nonlocality and communication complexity, Rev. Mod. Phys. \textbf{82}, 665 (2010).
\bibitem{a8a} R. Colbeck, and R. Renner, Free randomness can be amplified, Nat. Phys. \textbf{8}, 450 (2012).
\bibitem{a9a} C. Dhara, G. Prettico, and A. Antonio, Maximal quantum randomness in Bell tests, Phys. Rev. A \textbf{88}, 052116 (2013).
\bibitem{a10a} M. H. Li, X. J. Zhang, W. Z. Liu, et al., Experimental Realization of Device-Independent Quantum Randomness Expansion, Phys. Rev. Lett. \textbf{126}, 050503 (2021).
\bibitem{a11a} W. Z. Liu, M. H. Li, S. Ragy, et al., Device-independent randomness expansion against quantum side information, Nat. Phys. \textbf{17}, 448 (2021).
\bibitem{a12a} L. K. Shalm, Y. B. Zhang, J. C. Bienfang, et al., Device-independent randomness expansion with entangled photons, Nat. Phys. \textbf{17}, 452 (2021).
\bibitem{a13a} G. Svetlichny, Distinguishing three-body from two-body nonseparability by a Bell-type inequality, Phys. Rev. D \textbf{35}, 3066 (1987).
\bibitem{a14a} Q. Y. Pan, J. L. Jing, Hawking radiation, entanglement, and teleportation in the background of an asymptotically flat static black hole, Phys. Rev. D \textbf{78}, 065015 (2008).
\bibitem{a15a} W. C. Qiang, G. H. Sun, Q. Dong, and S. H. Dong, Genuine multipartite concurrence for entanglement of Dirac fields in noninertial frames, Phys. Rev. A \textbf{98}, 022320 (2018).
\bibitem{a16a} S. Xu, X. K. Song, J. D. Shi, and L. Ye, Probing the quantum correlation and Bell non-locality for Dirac particles with Hawking effect in the background of Schwarzschild black hole, Phys. Lett. B \textbf{733}, 1  (2014).
\bibitem{a18a} N. Iizuka, and D. Kabat, Mutual information in Hawking radiation, Phys. Rev. D \textbf{88} 084010 (2013).
\bibitem{a19a} X. Liu, Z. Tian, J. Wang, and J. Jing, Radiative process of two entanglement atoms in de Sitter spacetime, Phys. Rev. D \textbf{97}, 105030 (2018).
\bibitem{a20a} E. Martín-Martínez, and I. Fuentes, Redistribution of particle and antiparticle entanglement in noninertial frames, Phys. Rev. A \textbf{83}, 052306 (2011).
\bibitem{a22a} J. C. Wang, Z. H. Tian, J. L. Jing, and H. Fan, Irreversible degradation of quantum coherence under relativistic motion, Phys. Rev. A \textbf{93}, 062105 (2016).
\bibitem{a21a} S. M. Wu, Z. C. Li, and H. S. Zeng, Quantum coherence of multipartite W-state in a Schwarzschild spacetime, EPL \textbf{129}, 40002 (2020).
\bibitem{a23a} S.W. Hawking, Breakdown of predictability in gravitational collapse, Phys. Rev. D \textbf{14}, 2460 (1976).
\bibitem{a24a} Y. Dai, Z. Shen, and Y. Shi, Quantum entanglement in three accelerating qubits coupled to scalar fields, Phys. Rev. D \textbf{94}, 025012 (2016).

\bibitem{a26a} J. He, S. Xu, and L. Ye, Measurement-induced-nonlocality for Dirac particles in Garfinkle–Horowitz–Strominger dilation space–time, Phys. Lett. B \textbf{756}, 278 (2016).
\bibitem{a27a} J. C. Wang, H. X. Cao, J. L. Jing, and H. Fan, Gaussian quantum steering and its asymmetry in curved spacetime, Phys. Rev. D \textbf{93}, 125011
(2016).
\bibitem{a25a} A. J. Torres-Arenas, Q. Dong, G. H. Sun, W. C. Qiang, and S. H. Dong, Entanglement measures of W-state in noninertial frames,
Phys. Lett. B \textbf{789}, 93 (2019).
\bibitem{a29a} J. C. Wang,  J. Deng, and J. L. Jing, Classical correlation and quantum discord sharing of Dirac fields in noninertial frames, Phys. Rev. A \textbf{81}, 052120 (2010).
\bibitem{a30a} Y. Yao,  X. Xiao, L. Ge,  X. G. Wang, and C. P. Sun, Quantum Fisher information in noninertial frames, Phys. Rev. A \textbf{89}, 042336 (2014).
\bibitem{a32a} J. C. Wang, and J. L. Jing, Multipartite entanglement of fermionic systems in noninertial frames, Phys. Rev. A \textbf{83}, 022314 (2011).
\bibitem{a34a} S. M. Wu, and H. S. Zeng, Genuine tripartite nonlocality and entanglement in curved spacetime, Eur. Phys. J. C \textbf{82}, 4 (2022). 
\bibitem{a33a}  T. G. Zhang, X. Wang, and S. M. Fei, Hawking effect can generate physically inaccessible genuine tripartite nonlocality, Eur. Phys. J. C \textbf{83}, 607 (2023). 
\bibitem{aa33aa} L. J. Li, F. Ming, X. K. Song, L. Ye, and D. Wang, Quantumness and entropic uncertainty in curved space-time, Eur. Phys. J. C \textbf{82}, 726 (2022).
\bibitem{aa34aa} S. M. Wu, H. S. Zeng, and T. Liu, Genuine multipartite entanglement subject to the Unruh and anti-Unruh effects, New J. Phys. \textbf{24}, 073004 (2022).
\bibitem{a35a}  D. Giulini, E. Joos, C. Kiefer, J. Kupsch,  I. O. Stamatescu, and H. D. Zeh, Decoherence and the Appearance of a Classical World in Quantum Theory. Springer, Berlin (1996).
\bibitem{a36a}  M. A. Schlosshauer, Decoherence and the Quantum to Classical Transition. Springer, Berlin (2007).
\bibitem{a37a}  J. Wang, and  J. Jing, Multipartite entanglement of fermionic systems in noninertial frames, Phys. Rev. A
\textbf{83}, 022314 (2011).
\bibitem{a38a}  S. KHAN, Entanglement of tripartite states with decoherence in non-inertial frames. Journal of Modern Optics, \textbf{59}, 250 (2012).
\bibitem{a39a}  S. M. Wu, Z. C. Li, and H. S. Zeng, Multipartite coherence andmonogamy relationship under the Unruh effect in an open system, Quantum Inf. Process, \textbf{20}, 277 (2021).
\bibitem{a40a}  T. G. Zhang, Y. Hong, and S. M. Fei, System–environment dynamics of GHZ-like states in
noninertial frames, al frames, Quantum Inf. Process, \textbf{22}, 331 (2023).
\bibitem{aa40aa} S. Haddadi, M. A. Yurischev, M. Y. Abd-Rabbou, M. Azizi, M. R. Pourkarimi and M. Ghominejad, Quantumness near a Schwarzschild black hole, Eur. Phys. J. C \textbf{84}, 42 (2024).
\bibitem{a42a}  D. E. Chang,  A. S. S{\o}rensen, P. R. Hemmer,  M. D. Lukin, Quantum optics with surface plasmons, Phys. Rev. Lett. \textbf{97}, 053002 (2006).
\bibitem{a43a}  A. F. Van Loo,  A. Fedorov,  K. Lalumie\`{r}e,  B. C. Sanders, A. Blais, and  A. Wallraff, Photon-mediated interactions between distant artificial atoms, Science \textbf{342}, 1494 (2013).
\bibitem{a44a} M. Siomau, and Ali A. Kamli, Defeating entanglement sudden death by a single local filtering, Phys. Rev. A \textbf{86}, 032304 (2012).
\bibitem{a46a} N. Metwally, Quantum filtering of accelerated qubit-qutrit system, Optik \textbf{178}, 524 (2019).
\bibitem{a47a} A. Rodriguez-Blanco, K. B. Whaley, and A. Bermudez, Suppressing amplitude damping in trapped ions: Discrete weak measurements for a nonunitary probabilistic noise filter, Phys. Rev. A \textbf{107}, 052409 (2023).
\bibitem{a48a} W. Y. Sun, D. Wang, J. Yang, and L. Ye, Enhancement of multipartite entanglement in an open system under non-inertial frames, Quantum Inf. Process, \textbf{16}, 90 (2017). 
\bibitem{aa48aa} W. Y. Sun, D. Wang, B. L. Fang, J. D. Shi, and L. Ye, The enhancement of quantum entanglement under an open Dirac
system with the Hawking effect in Schwarzschild space-time, Laser Phys. Lett. \textbf{15}, 065210 (2018).
\bibitem{a49a} R. Gallego, L.E. Wrflinger, A. Ac\'{\i}n, and M. Navascus, Operational Framework for Nonlocality, Phys. Rev. Lett. \textbf{109}, 070401 (2012).
\bibitem{a50a} J. D. Bancal, J. Barrett, N. Gisin, and S. Pironio, Definitions of multipartite nonlocality, Phys. Rev. A \textbf{88}, 014102 (2013).
\bibitem{a51a} K. Mukherjee, B. Paul, and D. Sarkar, Efficient test to demonstrate genuine three particle nonlocality, J. Phys. A Math. Theor. \textbf{46}, 465302 (2015).
\bibitem{a52a} K. Wang, Y. Liang, and Z. J. Zheng, Genuine tripartite nonlocality of GHZ state in noninertial frames, Quantum Inf. Process. \textbf{19}, 140 (2020).
\bibitem{a53a} K. Wang, and Z. J. Zheng, Violation of Svetlichny inequality in Triple Jaynes-Cummings Models, Sci. Rep. \textbf{10}, 6621 (2020).
\bibitem{a54a} Z. H. Ma, Z. H. Chen, J. L. Chen, C. Spengler, A. Gabriel, and M. Huber, Measure of genuine multipartite entanglement with computable lower bounds,
Phys. Rev. A \textbf{83}, 062325 (2011).
\bibitem{a55a} W. C. Qiang, G. H. Sun, Q. Dong, and S. H. Dong, Genuine multipartite concurrence for entanglement of Dirac fields in noninertial frames, Phys. Rev. A \textbf{98}, 022320 (2018).
\bibitem{a56a} S. M. Hashemi Rafsanjani, M. Huber, C. J. Broadbent, and J. H. Eberly, Genuinely multipartite concurrence of N-qubit X matrices, Phys. Rev. A \textbf{86}, 062303 (2012).
\bibitem{a57a} D. R. Brill, and J. A. Wheeler, Interaction of Neutrinos and Gravitational Fields, Rev. Mod. Phys. \textbf{29}, 465 (1957).
\bibitem{a58a} T. Damoar, and R. Ruffini, Black-hole evaporation in the Klein-Sauter-Heisenberg-Euler formalism, Phys. Rev. D \textbf{14}, 332 (1976).
\bibitem{a59a} S. M. Barnett, and P. M. Radmore, Methods in Theoretical Quantum Optics, Oxford University Press, New York, 67–80 1997.
\bibitem{a60a} R. Kerner, and R. B. Mann, Tunnelling, temperature, and Taub-NUT black holes, Phys. Rev. D \textbf{73}, 104010 (2006).
\bibitem{a61a} M. A. Nielsen and I. L. Chuang, Quantum Computation and Quantum Information (Cambridge University Press, Cambridge, (2000).
\bibitem{a62a} M. Siomau, and Ali A. Kamli, Defeating entanglement sudden death by a single local filtering, Phys. Rev. A \textbf{86}, 032304 (2012). 
\end{thebibliography}
\end{document}